\documentclass[reprint, aps, onecolumn, amsmath, amssymb]{revtex4-2}

\usepackage[margin=2.6cm]{geometry}

\usepackage{hyperref}
\hypersetup{
    colorlinks = true,
    urlcolor   = blue,
    citecolor  = black,
}
\hypersetup{colorlinks,linkcolor={blue},citecolor={blue},urlcolor={blue}} 
\usepackage{graphicx}
\usepackage{dcolumn}
\usepackage{bm}
\usepackage[separate-uncertainty = true,multi-part-units=single]{siunitx}
\usepackage{mathtools}
\usepackage{gensymb}
\usepackage{amsmath}
\usepackage{physics}
\usepackage{stmaryrd}
\usepackage[nameinlink,noabbrev]{cleveref}
\crefformat{section}{\S#2#1#3}%
\crefformat{subsection}{\S#2#1#3}
\crefformat{subsubsection}{\S#2#1#3}
\crefformat{figure}{#2Fig.~#1#3}
\crefmultiformat{figure}{Figs.~#2#1#3}{ and~#2#1#3}{, #2#1#3}{ and~#2#1#3}
\usepackage{amssymb}
\usepackage[svgnames]{xcolor}
\usepackage{wasysym}

\begin{document}

\preprint{APS/123-QED}

\title{Stable bubble formations in a depth-perturbed Hele-Shaw channel\\}

\author{Jack Lawless}
\author{Andrew L. Hazel}%
\author{Anne Juel}
\email{anne.juel@manchester.ac.uk}
\affiliation{Manchester Centre for Nonlinear Dynamics, The University of Manchester, Oxford Road, Manchester, M13 9PL, UK}
\author{Jack Keeler}
\affiliation{School of Mathematics, The University of East Anglia, Norwich Research Park, Norwich, NR4 7TJ, UK\\}

\date{\today}
             
\begin{abstract}
Deformable bubbles propagated by the flow of a viscous liquid in a planar Hele-Shaw channel of uniform depth tend to travel steadily along the channel's streamwise axis and pairs of neighbouring bubbles will either separate or coalesce because an individual bubble's propagation speed increases monotonically with its size. Thus, any group of bubbles will eventually rearrange itself in order of decreasing size and all of the bubbles will separate. We show that, by introducing a small geometric perturbation to the channel in the form of an axially-uniform depth reduction along its centreline, the system supports a multitude of stable bubble formations and this can disrupt the usual reordering by bubble size. The constituent bubbles of a stable formation lie in alternation on opposite sides of the depth-perturbation, retain fixed shapes and propagate steadily at the same speed. A stable formation is always led by the smallest of its constituent bubbles, which would be the slowest bubble in isolation. The leading bubble propagates as if it were isolated and the trailing bubbles reduce their speeds by adjusting their shapes and overlaps of the depth-perturbation in the perturbation fields of their preceding nearest-neighbours in order to match that of the leading bubble. The trailing bubbles can be arranged in any order and, hence, the number of stable formations increases factorially as the number of bubbles is increased. 
\end{abstract}

\maketitle

\section{Introduction}
\label{sec:introduction}
Dispersed two-phase flows are found in a variety of natural and industrial processes. The dispersed phase may consist of gas bubbles or liquid droplets that are suspended within a continuous liquid phase; examples include oceanic flows \citep{oceanic}, volcanic flows \citep{volcanic} and oil recovery processes \citep{oil_recovery}. The underlying complexities of such flows often leads to disorder (i.e. apparent randomness) of the dispersed phase. However, a number of dispersed flows have in fact been found to self-organise and form coherent spatial structures in confined geometries. This behaviour has been observed, for instance, in microfluidics, where novel techniques such as flow-focusing are used to generate large quantities of droplets or bubbles \citep{Anna}. For example, one-dimensional trains of monodisperse microdroplets were found to exhibit unsteady collective dynamics in the form of longitudinal and transverse vibrational modes in a quasi-two-dimensional channel through hydrodynamic interactions between the dipolar flow fields that are induced by neighbouring droplets \citep{beatus, schiller, beatus_2}. The introduction of controlled geometrical perturbations to the confining channel has also been found to promote self-organisation in the flow of microdroplets. For example, upon encountering a channel expansion, one-dimensional trains of double emulsion microdroplets formed regularly spaced clusters through coalescence of their coating films \citep{kerstin}. Furthermore, clusters of microdroplets were found to spontaneously rearrange and assemble into a variety of complex `building block'-like structures due to the coupling between inter-droplet hydrodynamic interactions and adhesive depletion forces \citep{shen}. The introduction of an axial gradient in the channel's depth profile was found to promote the self-organisation of microdroplets into ordered two- and three-dimensional arrays \citep{parthiban}. Finally, self-organisation can also be achieved via external actuation. For example, groups of microbubbles were found to form periodic crystal-like lattices when excited by an acoustic field through the interacting surface waves that are emitted by the bubbles \citep{rabaud}. Importantly, the bubbles or droplets exhibit little to no deformation in each of the aforementioned examples because they are sufficiently small such that surface tension has the dominant influence on the interface shapes. However, in this paper, we focus on the self-organisation of larger bubbles, for which the restoring effects of surface tension cannot entirely counteract the deformation that is induced by the surrounding liquid flow.
 
\begin{figure}
\includegraphics[width=\textwidth, clip]{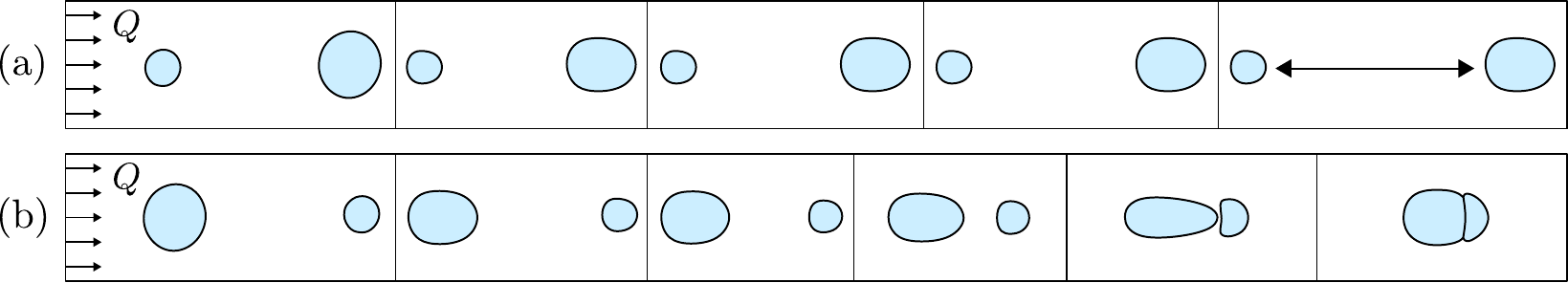}
\caption{The typical dynamics of a pair of initially well-separated air bubbles in a horizontally levelled planar Hele-Shaw channel of uniform depth when they are subjected to a constant volume flux flow of a surrounding viscous liquid. Time increases from left-to-right. (a) The larger bubble is initially situated in the leading position and the two bubbles separate indefinitely. (b) The smaller bubble is initially situated in the leading position and the two bubbles aggregate. These experiments were performed in the set-up that is described in Sec.~\ref{sec:setup} without a depth perturbation.}
\label{fig:uniform_channel}
\end{figure}

An idealised confined geometry to study multi-bubble dynamics is that of a Hele-Shaw channel. A Hele-Shaw channel consists of two parallel plates that are separated by a small gap distance $H^*$ relative to their width $W^*$. In this particular geometry, the interface that separates the two phases has a propensity to deform if the propagating bubbles are sufficiently large. Even in the absence of inertia, at finite surface tension numerical simulation or experiment is required to determine the bubble width (speed) that results from a combination of viscous and surface tension effects, whose ratio is quantified by the capillary number $\mathrm{Ca} = \mu U_b^* / \sigma$, where $\mu$ is the dynamic viscosity of the liquid, $\sigma$ is the surface tension and $U_b^*$ is the bubble's propagation speed; see, e.g., \cite{mclean1981}. 
In general, pairs of bubbles in a Hele-Shaw channel of uniform depth either separate or coalesce. The terminal rising speed of a free-rising bubble in an inclined Hele-Shaw channel increases monotonically with its size at low values of the Reynolds number $\mathrm{Re} = \rho U_b^* W^* / \mu$, where $\rho$ is the density of the liquid, and, thus, a pair of co-axial bubbles that is led by the smallest bubble will typically aggregate. However, in the reverse configuration, the long-term outcome of the bubbles is determined by their initial separation: well-separated bubbles tend to separate indefinitely whereas aggregation may occur if their initial separation is sufficiently small. The combination of these two dynamical behaviours leads to the formation of steadily propagating compound bubble clusters, similar to the one shown in the final panel of Fig.~\ref{fig:uniform_channel}\textcolor{blue}{(b)}, that separate indefinitely \citep{maxworthy_1986}. However, the constituent bubbles of a cluster ultimately coalesce due to the slow drainage of the lubrication films that separate them \citep{drainage}. Although similar long-term dynamics are observed in the high-$\mathrm{Re}$ regime, the transient dynamics are considerably richer because free-rising bubbles generate unsteady wake flows in their paths. The core of the wake that is generated by the leading bubble vertically entrains the trailing bubble and regularly sheds counter-rotating vortices that lead to the periodic ejection and subsequent realignment of the trailing bubble with the core of the wake as it rises \citep{Huisman, filella_2015, filella}. Broadly similar dynamics were reported in studies of homogeneous swarms of free-rising bubbles dispersed throughout a channel \citep{bouche, ruiz-rus}. 

When pairs of bubbles are driven by the constant volume flux flow of a surrounding viscous liquid in a planar Hele-Shaw channel, they exhibit broadly similar long-term behaviours to free-rising bubbles. The propagation speed of an isolated bubble increases monotonically with its size and, thus, pairs of bubbles of different sizes will either separate indefinitely or aggregate [Fig.~\ref{fig:uniform_channel}]. However, alternative forms of collective long-term behaviours have been uncovered in analytical studies of multiple bubbles propagating in a Hele-Shaw channel for extreme values of the surface tension. For example, in the limit of infinite surface tension (i.e. in the limit $\mathrm{Ca} \rightarrow 0$), one-dimensional trains of small and circular bubbles of identical size were found to exhibit longitudinal vibrational motion when driven by the flow of a surrounding viscous liquid \citep{booth}. The vibrational propagation modes were found to give rise to collective dynamics likened to that of a Newton's cradle through the aggregation and subsequent breakup of neighbouring bubble pairs. Conversely, in the absence of surface tension (i.e. in the limit $\mathrm{Ca} \rightarrow \infty$), groups of deformable bubbles were found to propagate steadily in a variety of fixed spatial arrangements when driven by the flow of a surrounding viscous liquid \citep{vasconcelos_2015}. The bubbles neither separate nor aggregate but, instead, maintain a constant separation from one another as they propagate. However, this study did not include a stability analysis and, thus, it is unknown if these arrangements could be observed in a real-world channel.

\begin{figure}
\includegraphics[width=\textwidth, clip]{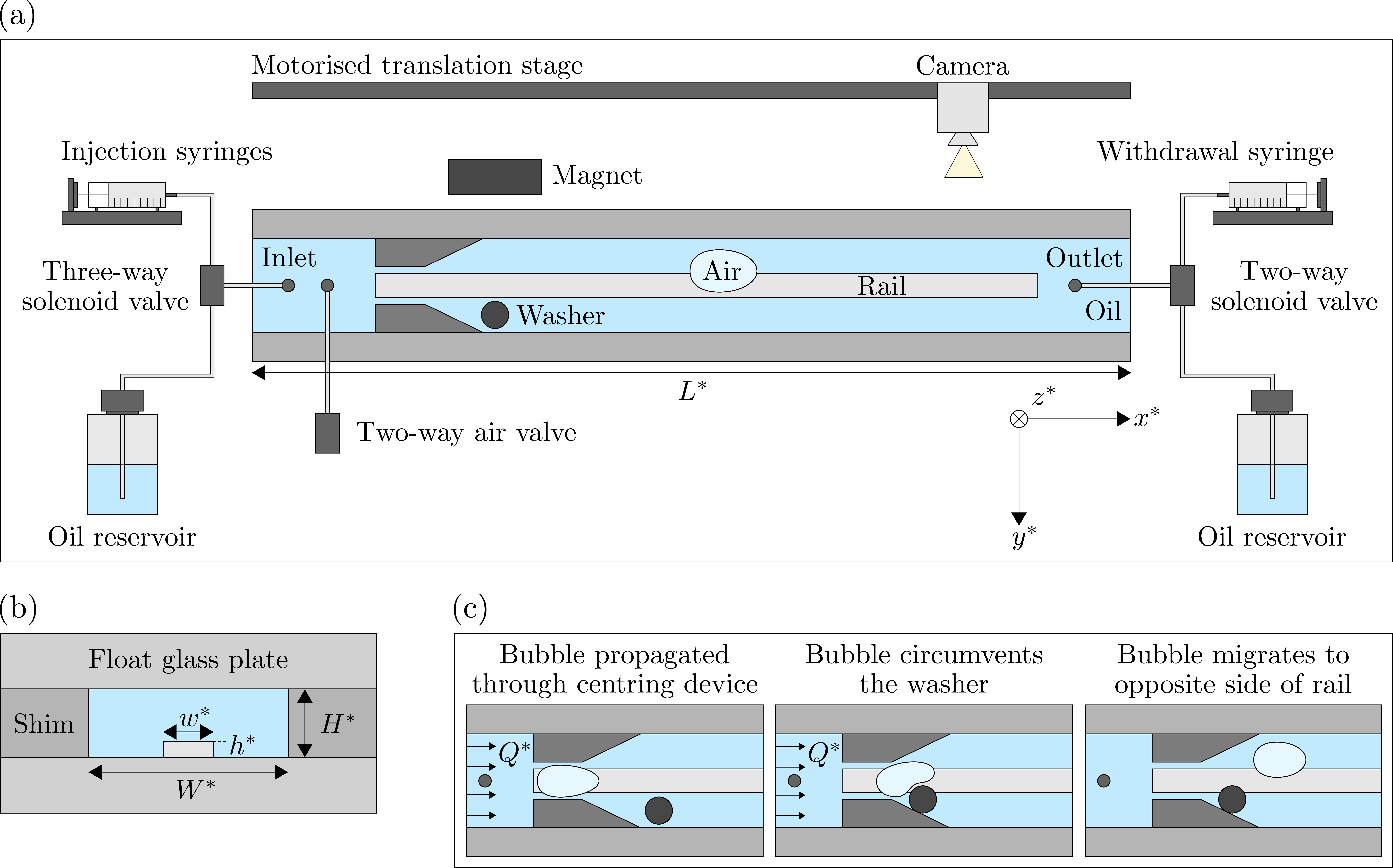}
\caption{(a) A schematic diagram of the Hele-Shaw channel that contains a small depth reduction or `rail' along its centreline. The bubbles are generated at the air injection port and are then guided to an asymmetric initial position by constricting the channel with a stainless steel washer. The syringe pumps inject oil into the channel inlet at a constant volumetric flux $Q^*$. (b) Cross-sectional schematic diagram of the channel ($W^* \gg H^*$). (C) A schematic diagram of the procedure that is used to initialise a bubble on a particular side of the rail.}
\label{fig:setup}
\end{figure} 

In this paper, we perform experiments at finite values of the capillary number and demonstrate how groups of bubbles of different sizes can self-organise into a multitude of stable formations in a Hele-Shaw channel that possesses a small non-uniformity in its geometry. The result is particularly striking because, to the best of our knowledge, 
only trains of equal-sized bubbles spaced far enough apart such that there are no interactions between them can be supported in a Hele-Shaw channel of uniform depth. The geometric non-uniformity is a small axially-uniform depth reduction along the centreline of the channel's cross-section, which is henceforth referred to as a rail. The Reynolds number remains small throughout our experiments and, hence, inertial forces are negligible. The system is nonlinear, however, because of the presence of air-fluid interfaces. 

The study of isolated bubble propagation in this system, both experimentally \citep{LifeAndFate, Lawless} and numerically \citep{franco_gomez_2018, Keeler2019}, has revealed a rich dynamical landscape that arises as a direct consequence of the variation in channel depth. For sufficiently large bubbles in an equivalent uniform channel, only a single solution branch is found to be stable at all values of the driving flow rate \citep{tanveer} and it corresponds to the steady, symmetric, single-tipped bubbles observed in experiments. In the channel with a rail, additional solution branches are stabilised due to the prescribed variation in the channel's depth. Notably, a steady asymmetric branch is found to be stable for all $\mathrm{Ca}$ \citep{Keeler2019, LifeAndFate}. Furthermore, in contrast to a Hele-Shaw channel of uniform depth, bubble breakup occurs more readily in the system with a rail, which motivated \citet{Keeler2021} to study the dynamics of two interacting bubbles. A family of stable two-bubble formations, characterised by the bubbles propagating steadily on opposite sides of the rail, were uncovered in their study through a combination of experiments and numerical simulations of a two-dimensional depth-averaged lubrication model. However, the detailed behaviour of these two-bubble states and the implications for steadily propagating states with more than two bubbles were not investigated.

In this paper, we establish the conditions for which stable two-bubble formations occur by identifying the key physical mechanisms at play. We then build on this understanding of stable two-bubble formations to establish a set of ``design principles" for three-bubble formations and examine their behaviours in a response to changes in the system's parameters. We then generalise our results for formations with larger numbers of bubbles. The paper is organised as follows. The experimental set-up and protocols that are implemented in order to reproducibly generate bubbles of prescribed sizes are described in Sec.~\ref{sec:setup}. Explaining the origin of the multi-bubble states relies on a thorough understanding of the behaviour of individual bubbles and, thus, we benchmark the propagation of isolated bubbles in Sec.~\ref{sec:isolated}. An outline of the two-dimensional depth-averaged lubrication model and its numerical implementation is provided in Sec.~\ref{sec:numerics}. We investigate the range of existence of stable two-bubble formations and their behaviours in response to varying the bubble sizes, flow rate and rail height in Sec.~\ref{two_bubble}. The design principles governing the existence of larger stable formations of bubbles are established in Sec.~\ref{N_bubble}. We proceed to discuss the long-term behaviour of an arbitrary train of bubbles in Sec.~\ref{bubble_train} and, finally, conclusions are drawn in Sec.~\ref{discussion}.


\section{Methods}
\subsection{Experimental methods}
\subsubsection{Experimental set-up and protocols}
\label{sec:setup}

The experiments in this study were performed in the same Hele-Shaw channel that was described by \citet{LifeAndFate} who provide detailed descriptions of both the experimental set-up [Fig.~\ref{fig:setup}\textcolor{blue}{(a)}] and protocols that were implemented in order to reproducibly generate and propagate bubbles of prescribed sizes. Here, we will recall the salient details. The channel consisted of two rectangular float glass plates that were separated by two parallel strips of stainless steel shim [Fig.~\ref{fig:setup}\textcolor{blue}{(b)}]. The shim's thickness $H^* =$ \SI{1.00 (1)}{\milli\metre} was measured at uniformly spaced points along the channel's length with a micrometer screw gauge and they were separated by a distance $W^* =$ \SI{40.0 (1)}{\milli\metre} with a stainless steel gauge block. The rail was formed by applying a thin strip of translucent PET tape along the centreline of the lower glass plate. The width of the tape was $w^* = $ \SI{10.0 (1)}{\milli\metre}. We performed experiments with two different rail thicknesses $h^*$; the majority of experiments that are described in this paper had $h^* =$ \SI{24 (1)}{\micro\metre} whilst a smaller number of experiments had $h^* =$ \SI{10 (1)}{\micro\metre}. Thus, the effective height of the channel in the region above the rail was reduced by either $2.4\%$ or $1\%$. The channel was filled with silicone oil (Basildon Chemicals Ltd) of dynamic viscosity $\mu =$ \SI{0.019}{\pascal \second}, density $\rho =$ \SI{951}{\kg \per \cubic \metre} and surface tension $\sigma =$ \SI{21}{\milli \newton \per \metre} at the ambient laboratory temperature of \SI{21(1)}{\celsius}. The flow of silicone oil was controlled by a network of three syringe pumps that were connected in parallel. The syringe pumps, channel inlet and an external oil reservoir were connected by a three-way solenoid valve. The channel outlet and an external oil reservoir were connected by a two-way solenoid valve. Thus, oil could either be driven into the channel at a constant flow rate $Q^*$ or withdrawn from the external oil reservoir in order to refill the injection syringes. The injection syringes and solenoid valves were controlled by a custom-built LabVIEW script. 

We generated air bubbles at the air injection port that was situated a short distance downstream of the channel inlet by slowly withdrawing a prescribed volume of oil through a syringe that was connected to the channel outlet whilst the two-way air valve was open to the atmosphere. Once generated, the bubble was detached and propagated away from the air valve by imposing a small flow of oil. This process was repeated several times in succession in order to generate multiple bubbles of prescribed volumes. The initial separation between the bubbles was controlled by varying the amount of oil that was injected following detachment. The bubbles were then propagated through a `centring device', which consisted of a symmetric channel constriction and a linear expansion region. Upon exiting the centring device, each of the bubbles were guided to an asymmetric initial position by applying an asymmetric constriction to the channel with a magnetic washer [Fig.~\ref{fig:setup}\textcolor{blue}{(c)}]. The position of the washer was controlled non-invasively by an N52-grade neodymium magnet that was situated underneath the lower glass plate and, by varying the side on which the constriction was applied, the bubbles could be initialised asymmetrically on either side of the rail.

\begin{figure}
\includegraphics[width=\textwidth, clip]{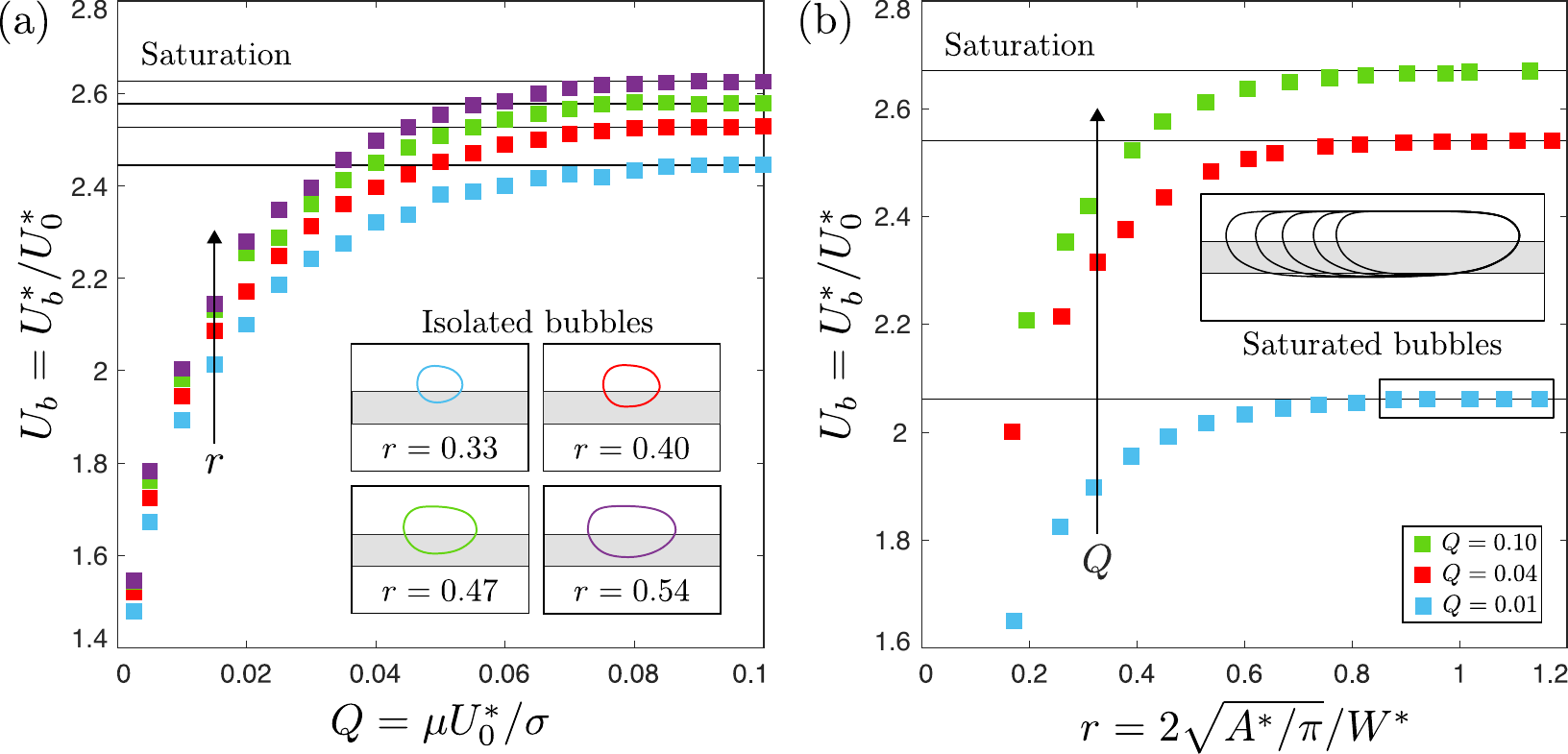}
\caption{(a) Variation of $U_b$ with $Q$ for isolated bubbles of various sizes. Inset: experimental snapshots of the bubbles at $Q=0.01$. (b) Variation of $U_b$ with $r$ at three values of $Q$ that span the range of investigated flow rates. Inset: superimposed experimental snapshots of saturated ($r \geq r_s$) bubbles at $Q=0.01$. The bubbles are aligned in the streamwise direction at their tips. The shape of the bubble's front and rear are unchanged for $r \geq r_s$. The rail's height is $h = h^*/H^* = 0.024$.}
\label{fig:single_bubble_speed}
\end{figure}

The bubbles were set into motion by imposing a constant flow rate $Q^*$ and filmed in top-view by an overhead CMOS camera, which was mounted onto a motorised translation stage. The camera was programmed to translate at a constant speed, which was selected via an empirical relationship between $U_b^*$ and $Q^*$ to ensure that the bubbles remained within its field of view. The observation window of the channel was $1760 \cross 330$ pixels and, depending on the value of $Q^*$, the camera's frame rate was varied between 20 and 60. The channel was uniformly illuminated from below by a custom-built LED light box, which led to the air-fluid interface appearing darkened due to light refraction. We used a Canny edge-detection algorithm in order to identify the bubble contours.

We will adopt the channel's half-width $W^*/2$ and the mean speed of oil in an equivalent uniform channel $U_0^* = Q^* / W^* H^*$ as our characteristic length and velocity scales. The imposed flow rate is parametrised in terms of the capillary number $Q = \mu U_0^* / \sigma$ that is based on the mean speed of the oil. The bubble's speed $U_b^*$ was determined by calculating the streamwise displacement of its centre of mass across a series of consecutive frames. The dimensionless bubble speed is $U_b = U_b^* / U_0^*$ and takes values between 1 and 3 in experiments. The capillary number in related literature is usually based on the bubble's speed, i.e. $\mathrm{Ca} = Q U_b$. However, in an attempt to avoid confusion, we will use $Q$ as the principal dynamic parameter throughout this paper. The bubble's in-plane area $A^*$ was determined from the Canny edge detection algorithm and its size is parametrised in terms of a dimensionless radius $r = 2 \sqrt{A^* / \pi} / W^*$. 

\subsubsection{Isolated bubbles}
\label{sec:isolated}
Throughout this paper, we will primarily work with four bubbles of size $r=0.33$, $r=0.40$, $r=0.47$ and $r=0.54$ and they will be represented by blue, red, green and purple-coloured contours, respectively. We begin by benchmarking the behaviour of each bubble when propagated in isolation. Here, results are shown for the specific asymmetric states that ``combine'' to form the multi-bubble states (i.e. stable formations) of interest. Although general information about these and other isolated-bubble states is given in \citep{LifeAndFate, Keeler2021}, the data presented here are new. Fig.~\ref{fig:single_bubble_speed}\textcolor{blue}{(a)} shows the dimensionless speed of each isolated bubble as a function of the dimensionless flow rate and, in each case, the bubble's speed increases sharply from the quasistatic (i.e. $Q=0$) limit and saturates for larger flow rates. We note that this behaviour is broadly similar to that of a semi-infinite air finger, which is the limiting case as the bubble's size tends to infinity. Fig.~\ref{fig:single_bubble_speed}\textcolor{blue}{(b)} shows the dimensionless speed of an isolated bubble as a function of its size for several values of $Q$. The bubble's dimensionless speed increases monotonically with its size from the tracer-like (i.e. $r=0$) limit, in which it moves at the fluid speed, and saturates at $r = r_s \approx 0.90$. For $r \geq r_s$, the bubble elongates in the streamwise direction whilst the in-plane curvature of its front and rear remain unchanged [inset of Fig.~\ref{fig:single_bubble_speed}\textcolor{blue}{(b)}]. The bubble's speed is determined by the local pressure gradient ahead of its front, which is affected by changes in the viscous dissipation that arise as it displaces the viscous fluid, as well as the surface tension-induced pressure jump at the interface. The details depend on both the shape and location of the bubble within the non-uniform channel and we are not aware of any simpler predictive models other than numerical solutions of the depth-averaged system that is described in the next subsection.

\begin{figure}
\includegraphics[width=\textwidth, clip]{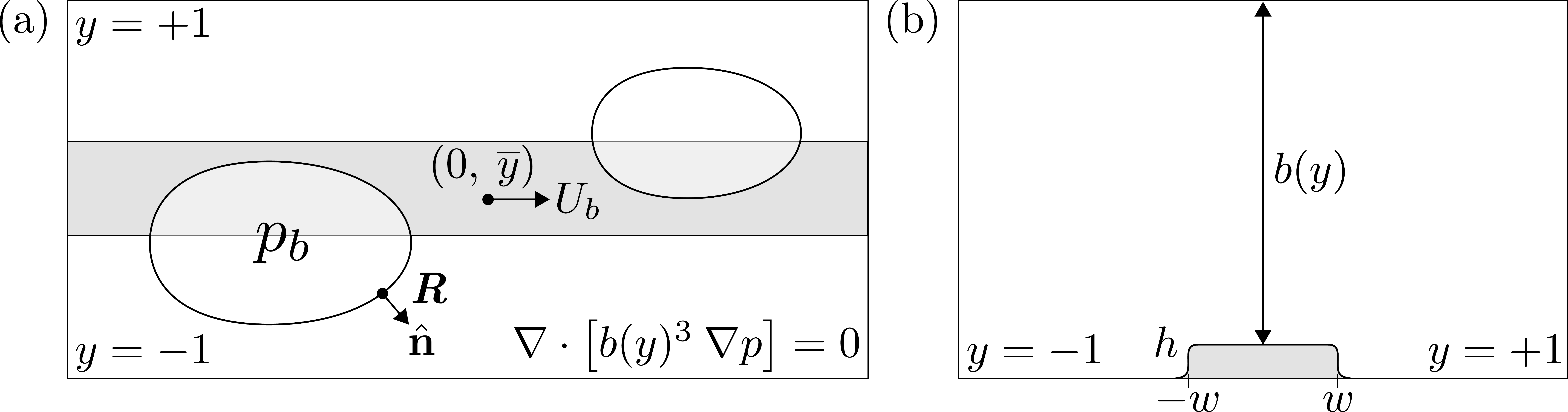}
\caption{(a) The truncated dimensionless computational domain. The streamwise component of the bubbles' centre of mass is constrained to remain fixed at zero. (b) Cross-sectional diagram of the computational domain. The variable channel depth is modelled by a piecewise-smooth depth profile $b(y)$ [Eq. \ref{rail}].}
\label{fig:numerical_schematic}
\end{figure}

\subsection{Mathematical model and numerical methods}
\label{sec:numerics}
We complement experiments with numerical simulations of a two-dimensional depth-averaged lubrication model whose validity has been confirmed in studies of isolated \citep{LifeAndFate} and multiple bubbles \citep{Keeler2021}. We opt to describe only the model's salient details because its general implementation is unchanged from previous studies.

\subsubsection{Model outline}
The truncated dimensionless computational domain and its cross-section are shown in Fig.~\ref{fig:numerical_schematic}. The rail is modelled by a piecewise-smooth depth profile $b(y)$ of the form
\begin{equation}
\label{rail}
    b(y) = 1-\frac{h}{2} \bigg{[} \tanh{\big{(}s(y+w) \big{)}} - \tanh{\big{(}s(y-w) \big{)}} \bigg{]},
\end{equation}
where $h = 0.024$ is the dimensionless height and $w = 0.25$ is the dimensionless width of the rail. The parameter $s = 40$ determines the sharpness of the rail's edges \citep{Thompson2014}.

We perform simulations in a co-moving frame of reference that translates with velocity $\textit{\textbf{U}}(t) = (U_b(t), \: 0)$, where $U_b(t)$ is an unknown of the problem that is obtained by requiring that the streamwise component of the centre of mass of the group of bubbles is fixed at zero. Upon applying lubrication theory \citep{reynolds} and depth-averaging, the three-dimensional Navier--Stokes equations in the fluid domain reduce to a single two-dimensional lubrication equation in terms of the dimensionless fluid pressure $p$,
\begin{equation}
    \grad \cdot \big{[} (b(y)^3 \: \grad{p} \big{]}=0.
\end{equation}
We impose no-penetration conditions on the channel side walls, i.e. $p_y = 0$ at $y= \pm 1$. We denote the coordinates of a point on the air-fluid interface by $\textbf{R}$ and impose a no-penetration condition:
\begin{equation}
    \label{eqn:kinematic}
    \frac{\partial \textbf{R}}{\partial t} \cdot \hat{\textbf{n}} = \big{[} -b(y)^2 \: \grad{p} - \textit{\textbf{U}}(t) \big{]} \cdot \hat{\textbf{n}},
\end{equation}
where $\hat{\textbf{n}}$ is the outward-pointing unit normal vector to the interface at $\textbf{R}$. The pressure jump across the interface, arising exclusively due to surface tension, is described by the Young--Laplace equation: 
\begin{equation}
\label{eqn:dynamic}
    \llbracket p \rrbracket ^{\textnormal{bubble}}_{\textnormal{fluid}} = \frac{1}{3 \alpha Q} \bigg{[} \frac{\kappa}{\alpha} + \frac{1}{b(y)} \bigg{]},
\end{equation}

\noindent where the dimensionless in-plane curvature of the interface at $\textbf{R}$ is denoted by $\kappa$. The transverse curvature term $1/b(y)$ arises under two simplifying assumptions: (i) the bubble occupies the entirety of the channel depth and (ii) the interface is semi-circular in the transverse direction, with radius $b(y)/2$. The (constant) internal pressure of each bubble is determined by requiring that their dimensionless volumes remain constant. The mathematical model does not incorporate the presence of thin films that are deposited on the upper and lower channel boundaries as the bubble propagates in experiments. However, as in previous studies, we find that the model captures all of the dynamics that are observed in experiments. There is a quantitative discrepancy in the values of $Q$ at which particular phenomena occur between experiments and numerical simulations. This discrepancy increases as $Q$ is increased; we believe that this is due to the corresponding increase in the thickness of the neglected fluid films. We will discuss this issue in Sec.~\ref{two_bubble}.

\begin{figure}
\includegraphics[width=\textwidth, clip]{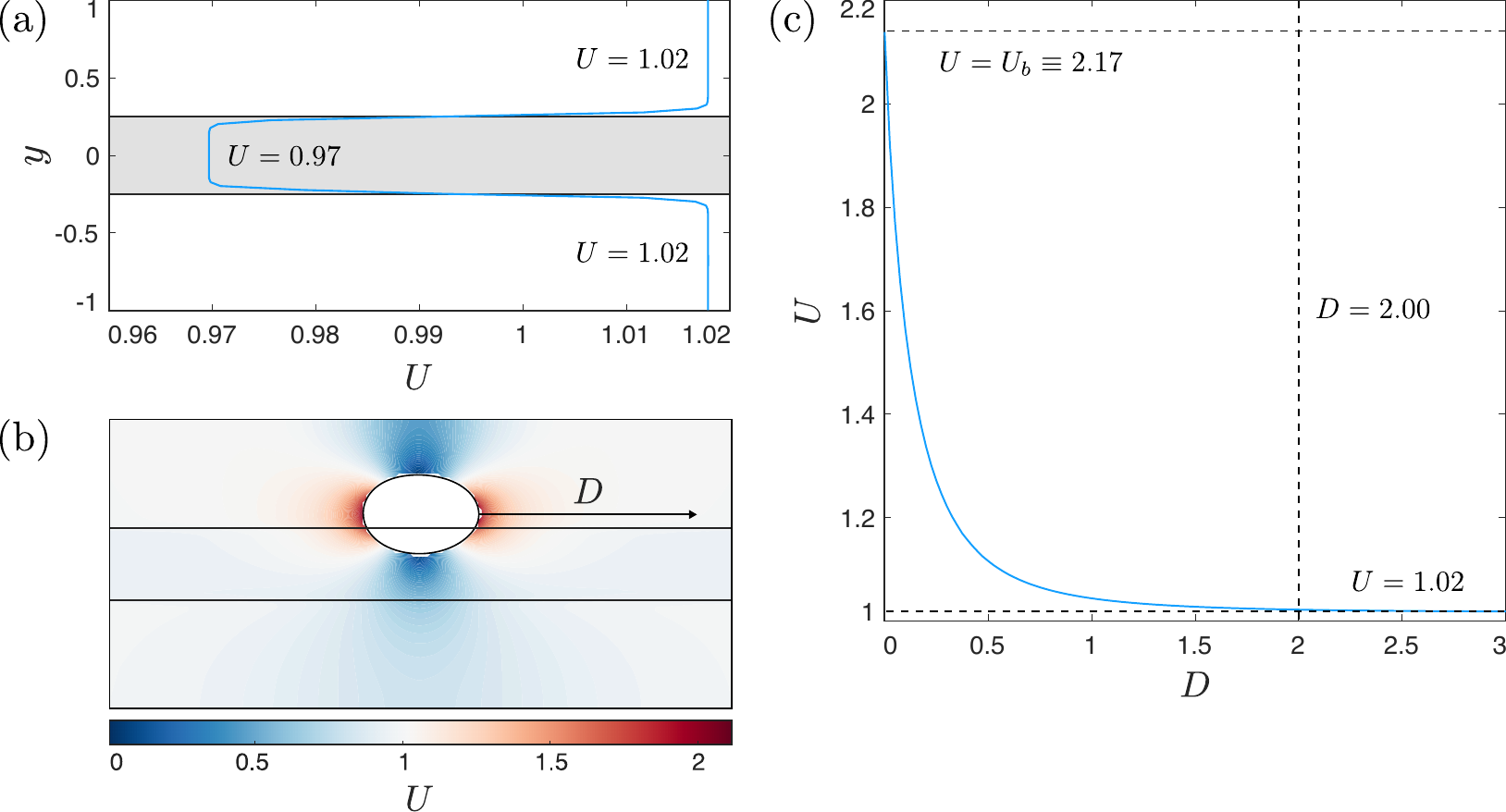}
\caption{(a) Variation of the dimensionless fluid speed $U = U^* / U_0^*$ with $y$ in a channel without a bubble at $Q=0.04$ in the depth-averaged lubrication model. (b) The flow field in the vicinity of an isolated bubble of size $r=0.33$ at $Q=0.04$ in the depth-averaged lubrication model. The dimensionless fluid speed is represented by the heat map. (c) Variation of $U = U^* / U_0^*$ with the dimensionless streamwise distance $D = 2 D^* / W^*$ from the bubble's tip.}
\label{fig:single_bubble_numerics}
\end{figure}

The system is spatially discretised by using a finite element method in the open-source library \texttt{oomph-lib} \citep{oomph}. We propagate elliptical bubbles from specified initial placements in the computational domain by imposing a constant pressure gradient in the streamwise direction. The value of the pressure gradient is chosen in order to obtain a specified dimensionless flow rate $Q$ of the surrounding fluid. The bubbles are initialised sufficiently far apart in order to prevent their immediate aggregation following the imposition of flow. The temporal evolution of the bubbles is simulated by employing a backwards difference time-stepping algorithm (BDF2) with a time-step $\Delta t = 0.01$. We have omitted details relating to the handling of the finite element mesh (see \citet{LifeAndFate}) but confirm that all of the presented numerical results are fully resolved.

\subsubsection{Isolated bubbles}
The depth-averaged lubrication model readily quantifies the flow field perturbations that are imposed by the asymmetrically propagating bubbles. Fig.~\ref{fig:single_bubble_numerics}\textcolor{blue}{(a)} shows the single-phase flow field $U = U(y)$. For this particular rail height, the fluid's dimensionless speed $U = U^* / U_0^*$ is given by $U = 0.97$ and $U = 1.02$ in the occluded and unoccluded regions, respectively, sufficiently far away from the rail's edges. The small decrease in the fluid's speed in the occluded region is a consequence of the local increase in the channel's viscous resistance.

The typical flow field perturbation that is imposed by the motion of an isolated bubble is presented in Fig.~\ref{fig:single_bubble_numerics}\textcolor{blue}{(b)}. The fluid's speed is represented by the heat map; blue-coloured regions correspond to $U < 1$ and red-coloured regions correspond to $U > 1$. The bubble's motion imposes a localised perturbation onto the surrounding flow field; its speed is more than twice that of the mean flow in the immediate vicinity of its front and rear whilst it decreases in the vicinity of its sides. The sharp decay of the flow field perturbation with increasing distance is evident in Fig.~\ref{fig:single_bubble_numerics}\textcolor{blue}{(b)}, where $U$ is plotted as a function of the dimensionless streamwise distance $D$ from the bubble's tip. The fluid's speed at the bubble's tip is equal to $U_b$ as a consequence of Eq.~\ref{eqn:kinematic} and the ambient fluid speed is reattained (within $0.5 \%$) at $D=2$. The decay rate of the flow field perturbation is slower than that predicted by simple dipole models that are used to model small bubbles in Hele-Shaw channels \citep{green} and this indicates that these bubbles are in a regime in which finite-size effects are important. 

\section{Results}
\label{sec:results}

\begin{figure}
\includegraphics[width=\textwidth, clip]{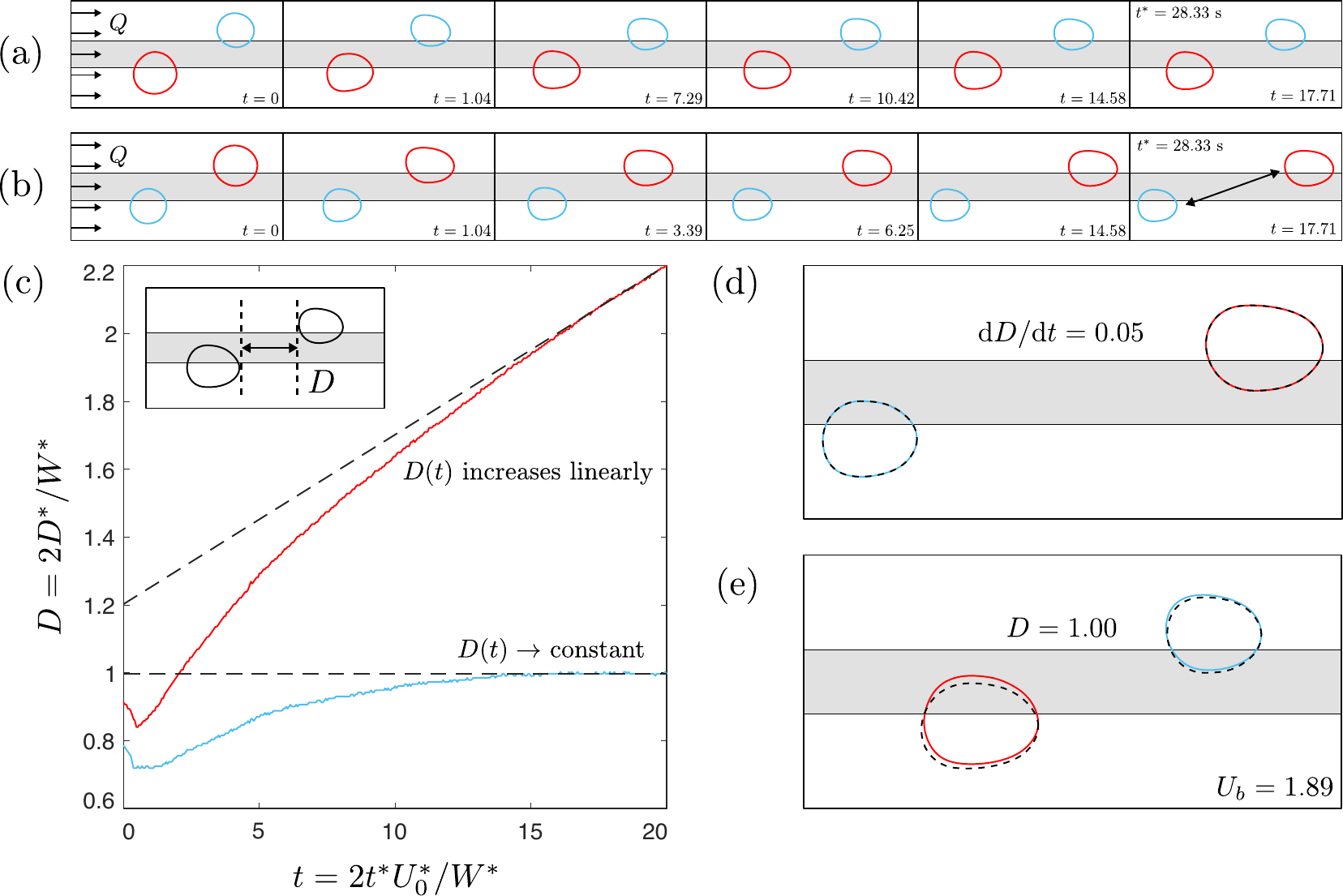}
\caption{(a, b) Experimental time-sequences of two bubbles of size $r=0.33$ and $r=0.40$ when propagated from rest at $Q=0.01$. The smaller bubble is initially situated in the leading position in (a), whereas the larger bubble is initially situated in the leading position in (b). (c) The corresponding time-evolutions of $D$. (d, e) Superposition of the final bubble shapes with their corresponding isolated bubbles (dashed black-coloured contours).}
\label{fig:two_bubble_state}
\end{figure}

\subsection{Stable two-bubble formations}
\label{two_bubble}
We begin by characterising the fundamental physical features of the stable formations of two bubbles that were identified by \citet{Keeler2021}. The experimental time-sequence in Fig.~\ref{fig:two_bubble_state}\textcolor{blue}{(a)} is a typical example of two bubbles organising into a stable formation when propagated from rest on opposite sides of the rail at $Q=0.01$; the leading bubble's size is $r_1 = 0.33$ and the trailing bubble's size is $r_2 = 0.40$. Thus, in the absence of any interaction, the trailing bubble would propagate faster than the leading bubble [see \S \ref{sec:isolated}]. Furthermore, if the two bubbles were located on the same side of the rail, or if there were no rail, the trailing bubble would catch up to the leading bubble and form an aggregate. However, the barrier between bubbles that is presented by the rail provides enough additional resistance to prevent their aggregation and, instead, the bubbles preserve their initial order following the imposition of flow and evolve transiently until their shapes and separation do not change with time. This behaviour is reflected in the temporal evolution of the dimensionless streamwise separation $D = 2D^* / W^*$ between the trailing bubble's front and the leading bubble's rear [Fig.~\ref{fig:two_bubble_state}\textcolor{blue}{(c)}]. The separation between the two bubbles in their stable formation is $D = 1.00$ and its speed $U_b = 1.89$ is equal to the leading bubble's isolated speed. However, upon reversing the order of the two bubbles in Fig.~\ref{fig:two_bubble_state}\textcolor{blue}{(b)}, because the trailing bubble propagates more slowly than the leading bubble, they proceed to separate indefinitely. The gradient of $D$ (i.e. the rate at which the two bubbles separate) decreases monotonically and tends towards a constant value $\textnormal{d}D / \textnormal{d} t = 0.05$ for $D > 2$ and this is equal to the difference between the speeds of the two corresponding steadily propagating isolated bubbles. 

We have superimposed the shapes of these isolated bubbles (dashed black-coloured contours) onto the final panels. The two separating bubbles adopt identical shapes to their isolated counterparts for $D > 2$ because they are effectively non-interacting [Fig.~\ref{fig:two_bubble_state}\textcolor{blue}{(d)}]. However, the two bubbles adopt different shapes in their stable formation [Fig.~\ref{fig:two_bubble_state}\textcolor{blue}{(e)}]. The shape of the leading bubble's front does not change but its rear inclines modestly away from the channel's centreline. By contrast, the trailing bubble changes its shape significantly because it broadens and increases its overlap of the rail. The changes in the shape and position of the trailing bubble are accompanied by a reduction in its speed to that of the leading bubble. The geometric remodelling of the trailing bubble is what allows the stable formation to exist.

\begin{figure}
\includegraphics[width=\textwidth, clip]{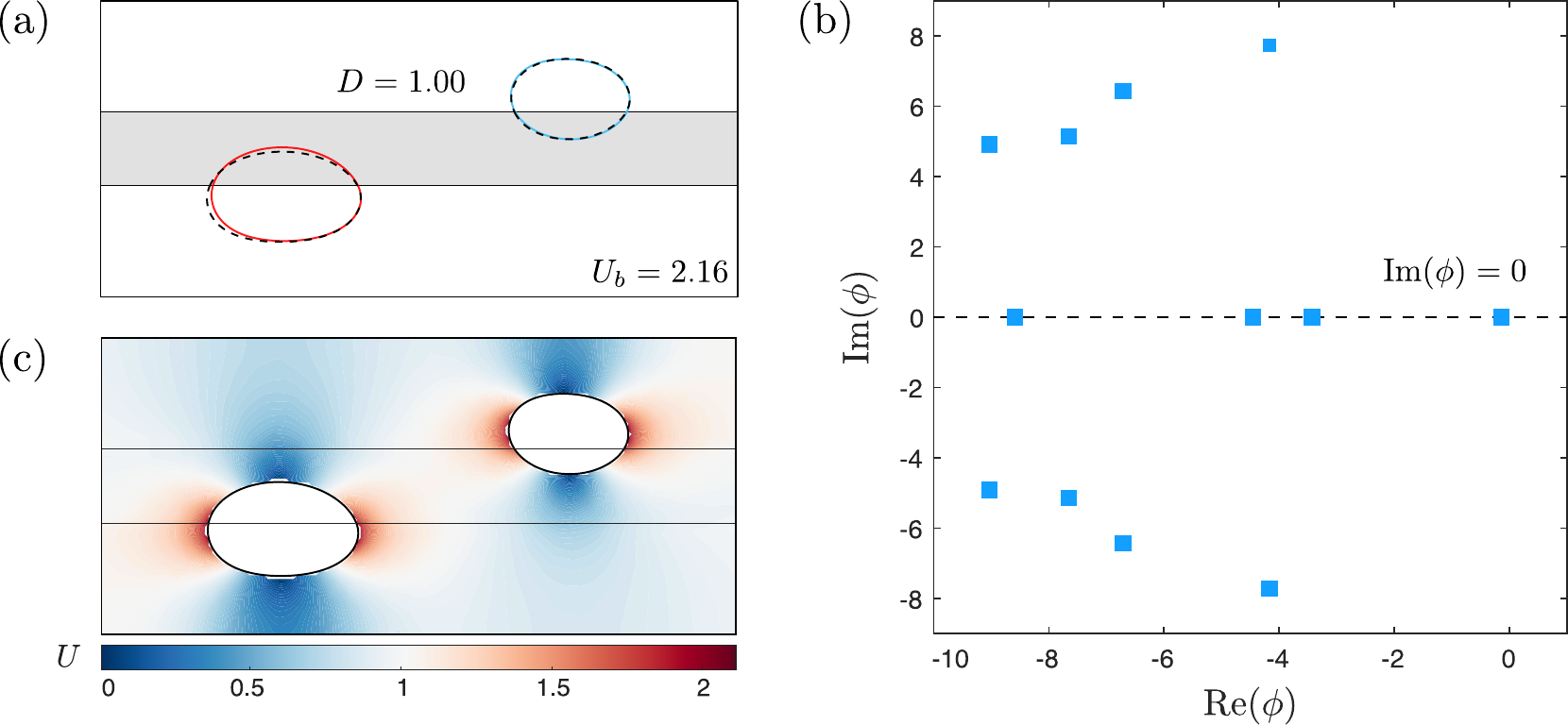}
\caption{(a) A stable two-bubble formation obtained from a numerical simulation of the depth-averaged lubrication model at $Q=0.04$. The two bubbles have sizes $r_{1}=0.33$ and $r_{2}=0.40$. (b) The formation's twelve least-stable eigenvalues. (c) A heat map of the dimensionless fluid speed $U$ in the vicinity of the bubbles.}
\label{fig:numerical_two_bubble}
\end{figure}

\begin{figure}
\includegraphics[width=\textwidth, clip]{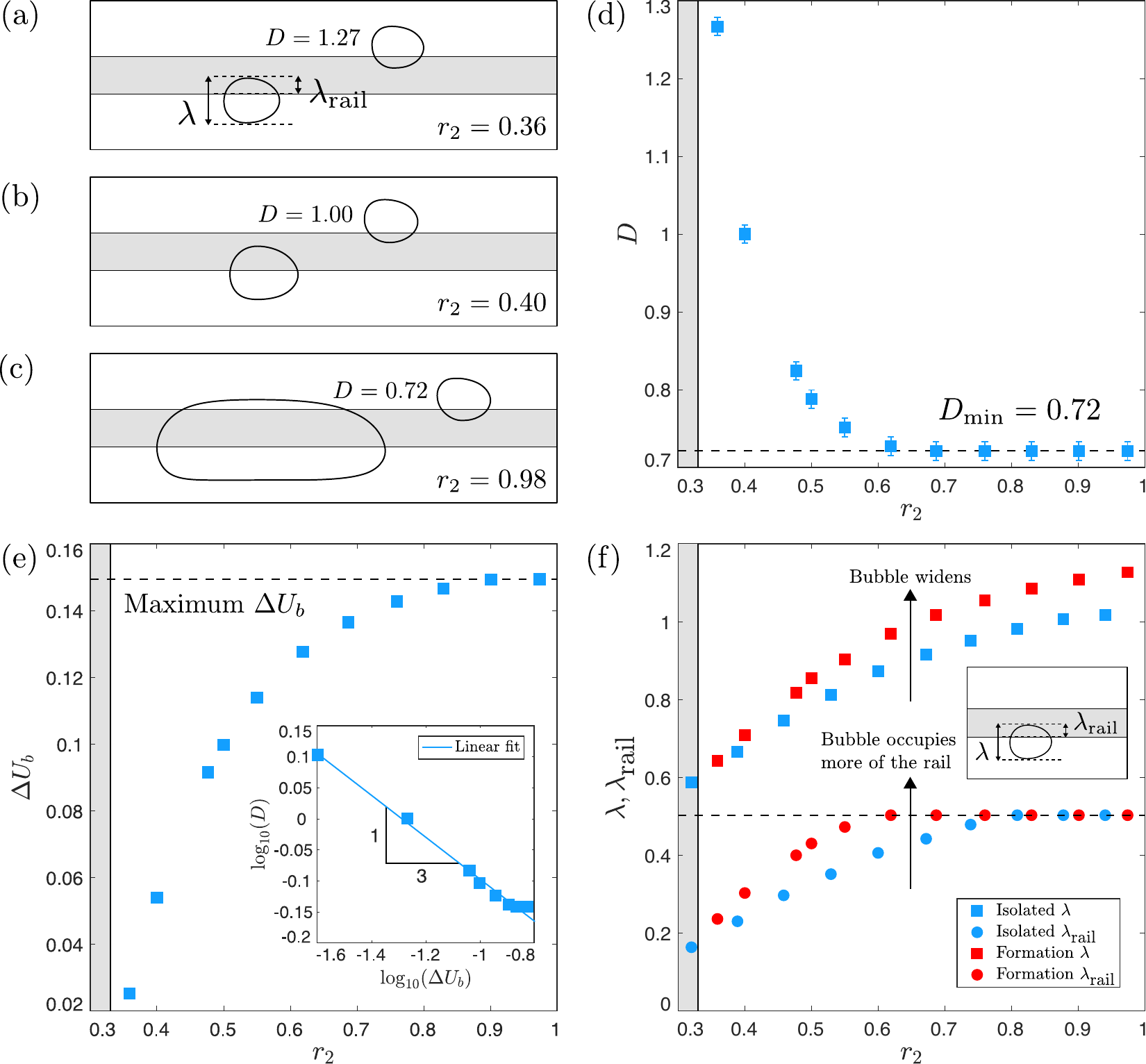}
\caption{(a--c) Experimental snapshots of stable two-bubble formations as the middle bubble's size $r_2$ is increased. The leading bubble's size $r_1 = 0.33$ and the dimensionless flow rate $Q=0.04$ are fixed. (d) Variation of $D = 2D^*/W^*$, the streamwise separation between the leading bubble's rear and the trailing bubble's front, with $r_2$. (e) Variation of $\Delta U_b$, the difference between the two isolated bubble speeds, with $r_2$. Inset: log-log plot of $D$ versus $\Delta U_b$. (f) Variation of $\lambda$, the maximum width of the trailing bubble's cross-section, and $\lambda_{\textnormal{rail}}$, the trailing bubble's maximum overlap of the rail, with $r_2$. Error bars have been omitted in (e) and (f) because they are smaller than the data markers.}
\label{fig:vary_r2}
\end{figure}

We find qualitatively similar behaviour in numerical simulations of the depth-averaged lubrication model. Fig.~\ref{fig:numerical_two_bubble}\textcolor{blue}{(a)} shows the numerically computed stable two-bubble formation for the same bubble sizes as in the experiment. However, we had to increase the dimensionless flow rate to $Q = 0.04$ in order to obtain the same separation ($D = 1.00$) between the bubbles. The increased flow rate is required because, for fixed flow rates, bubbles propagate at slower speeds in the numerical simulations compared to the experiments and this is a consequence of neglecting fluid films. The formation's twelve least-stable eigenvalues $\phi_i$ that were obtained from a linear stability analysis are plotted in Fig.~\ref{fig:numerical_two_bubble}\textcolor{blue}{(b)}: a combination of real, negative eigenvalues and complex conjugate eigenvalues with negative real components indicate that the state corresponds to a stable spiral in the system's phase space. The numerical model captures all of the qualitative characteristics that were identified in experiments. For example, the formation's speed $U_b = 2.16$ is marginally $(< 1 \%)$ slower than the leading bubble's isolated speed $U_b = 2.17$. We have superimposed the corresponding isolated bubble shapes (dashed black-coloured contours) onto the stable two-bubble formation in Fig.~\ref{fig:numerical_two_bubble}\textcolor{blue}{(a)}. The leading bubble's shape is visibly unchanged from that of its isolated counterpart, whereas the trailing bubble broadens and increases its overlap of the rail (albeit by a smaller amount than in experiments). The flow field ahead of the leading bubble is unaffected by the flow field perturbation that is imposed by the trailing bubble [Fig.~\ref{fig:numerical_two_bubble}\textcolor{blue}{(c)}]. However, the two bubbles form an adjoined flow field perturbation in the region between the leading bubble's rear and the trailing bubble's front because the bubbles are within their characteristic interaction range [Fig.~\ref{fig:single_bubble_numerics}\textcolor{blue}{(c)}]. We obtain a similar qualitative agreement for a variety of different flow rates and bubble sizes and, having thus established a broad agreement between experiments and the mathematical model, we will now proceed to explore the effects of key parameter variations on the behaviour of a stable two-bubble formation.

\subsubsection{Variation of $D$ with $r_2$}
\label{vary_r2}

The experimental snapshots in Figs.~\ref{fig:vary_r2}\textcolor{blue}{(a)}--\ref{fig:vary_r2}\textcolor{blue}{(c)} show the stable formation as the size of the trailing bubble $r_2$ is increased whilst the leading bubble's size $r_1=0.33$ and dimensionless flow rate $Q=0.04$ are fixed. Remarkably, the leading bubble's shape is hardly affected by the change in the trailing bubble's size. The shape of the leading bubble's front does not change because this region of the flow field is unaffected by the flow field perturbation that is imposed by the trailing bubble [Fig.~\ref{fig:numerical_two_bubble}\textcolor{blue}{(c)}]. However, its rear slightly inclines away from the channel's centreline as the trailing bubble's size is increased. The slight inclination of the leading bubble's rear does not change its speed and, thus, the stable formation's speed ($U_b = 1.89$) is constant for all values of $r_2$. 

The separation between the bubbles decreases monotonically as the trailing bubble's size is increased [Fig.~\ref{fig:vary_r2}\textcolor{blue}{(d)}] because the trailing bubble's isolated speed increases [Fig.~\ref{fig:single_bubble_speed}\textcolor{blue}{(b)}]. Thus, the reduction in the trailing bubble's speed $\Delta U_b$ that is required in order for it to match the leading bubble's speed also increases monotonically with $r_2$ [Fig.~\ref{fig:vary_r2}\textcolor{blue}{(e)}]. The increasing speed reduction that is needed for the bubbles to propagate in a stable formation requires an increasingly significant perturbation to the trailing bubble's shape and this is achieved by it moving closer to the leading bubble's rear so that it experiences a local pressure perturbation of increased magnitude. The linear variation of $D$ with $\Delta U_b$ on a log-log scale indicates an approximate power law behaviour over a limited range; see [inset of Fig.~\ref{fig:vary_r2}\textcolor{blue}{(e)}]. Eventually, the separation between the bubbles saturates at a minimum value $D = D_{\textnormal{min}} \equiv 0.72$ for $r_2 \geq r_s$ because the speed of the trailing bubble reaches a maximum value (i.e. the limit of a semi-infinite air finger) and, thus, $\Delta U_b$ saturates at a maximum value $\Delta U_b = \Delta {U_b}_{\textnormal{max}} = 0.15$. The separation between the bubbles increases sharply in the limit $r_2 \rightarrow r_1$ because $\Delta U_b \rightarrow 0$: the bubbles will travel at the same speed when their sizes are equal. This is the degenerate case that is preserved in a Hele-Shaw channel of uniform depth.

\begin{figure}
\includegraphics[width=\textwidth, clip]{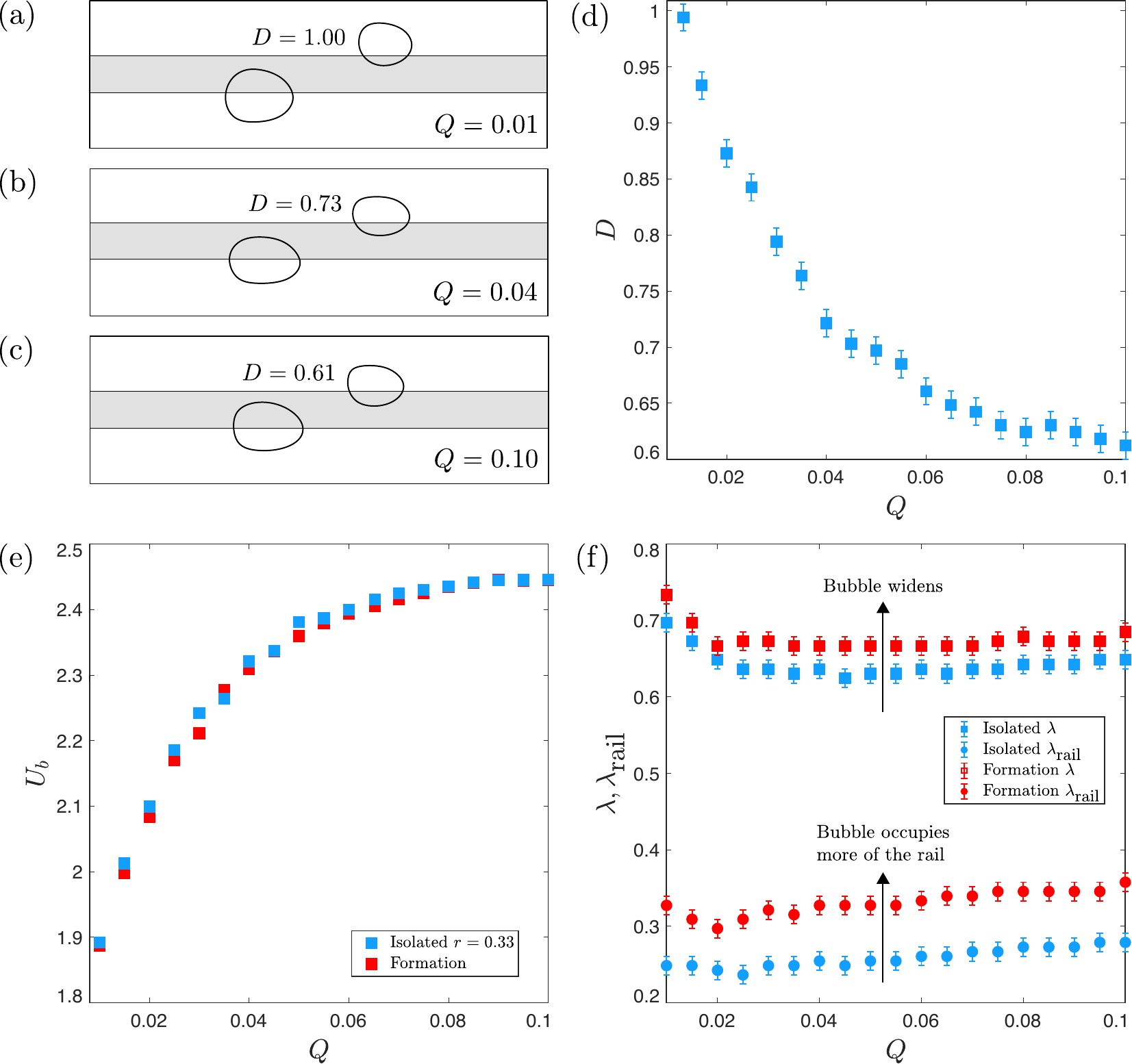}
\caption{(a--c) Experimental snapshots of a stable two-bubble formation as the flow rate $Q = \mu U_0^* / \sigma$ is increased. The leading bubble's size $r_1 = 0.33$ and the trailing bubble's size $r_2 = 0.40$ are fixed. (d) Variation of $D$, the streamwise separation between the leading bubble's rear and the trailing bubble's front, with $Q$. (e) Variation of $U_b$, the bubble's speed relative to the mean fluid speed, with $Q$. (f) Variation of $\lambda$, the maximum width of the trailing bubble's cross-section and $\lambda_{\textnormal{rail}}$, the trailing bubble's maximum overlap of the rail, with $Q$. Error bars have been omitted in (e) because they are smaller than the data markers.}
\label{fig:vary_Q}
\end{figure}

The trailing bubble's shape becomes increasingly perturbed (relative to its isolated state) as its size increases because a greater reduction in its speed is required in order for it to match the speed of the leading bubble. There are two different mechanisms by which reconfiguration of the trailing bubble can lead to a change in its speed: (i) its width can change and (ii) its position within the channel's cross-section can change. The maximum width of the trailing bubble's cross-section $\lambda = 2 \lambda^* / W^*$ and the portion of its cross-section that overlaps the rail $\lambda_{\textnormal{rail}} = 2 \lambda_{\textnormal{rail}} / W^*$ [defined in Fig.~\ref{fig:vary_r2}\textcolor{blue}{(a)}] deviate increasingly from their isolated counterparts as the middle bubble's size is increased [Fig.~\ref{fig:vary_r2}\textcolor{blue}{(f)}]. The trailing bubble fully overlaps the rail at $r_2=0.62$ (i.e. when $\lambda_{\textnormal{rail}} = 0.50$) and continues to broaden thereafter. However, the trailing bubble's speed continues to decrease after it fully overlaps the rail. This behaviour indicates that the reduction in the trailing bubble's speed is achieved by a complex interplay between the variation in its overlap of the rail (i.e. position within the channel's cross-section) and width (i.e. shape).

\subsubsection{Variation of $D$ with $Q$}
The experimental snapshots in Figs.~\ref{fig:vary_Q}\textcolor{blue}{(a)}--\ref{fig:vary_Q}\textcolor{blue}{(c)} show the stable formation as the dimensionless flow rate $Q$ is increased whilst the leading bubble's size $r_1=0.33$ and trailing bubble's size $r_2=0.40$ are fixed. The increased flow rate causes the bubbles to become more slender because of the increasing influence of viscous forces.

The separation between the bubbles decreases monotonically as the dimensionless flow rate is increased from the quasistatic ($Q=0$) limit [Fig.~\ref{fig:vary_Q}\textcolor{blue}{(d)}]. This behaviour, although qualitatively similar, is more complex than that which was described in Sec.~\ref{vary_r2} because the leading bubble's isolated speed (and, therefore, the formation's speed) increase [Fig.~\ref{fig:vary_Q}\textcolor{blue}{(e)}]. However, there is a negligible increase in the required reduction in the trailing bubble's speed $\Delta U_b$ as $Q$ is increased because the isolated speeds of both bubbles increase by approximately the same amount. Thus, the trailing bubble's maximum width and overlap of the rail both increase by an approximately constant value (i.e. the two corresponding curves are translated) as $Q$ is increased [Fig.~\ref{fig:vary_Q}\textcolor{blue}{(f)}] because the same reduction in speed requires a similar degree of geometric remodelling. We, therefore, attribute the decrease in the separation between the bubbles with increasing flow rate to the decreasing significance of the flow field perturbation that is induced by the leading bubble relative to the background pressure gradient because the trailing bubble needs to move closer to the leading bubble in order to experience the required flow field perturbation that results in the necessary geometric change.

\subsubsection{Two-bubble phase diagram}
\begin{figure}
\includegraphics[width=\textwidth, clip]{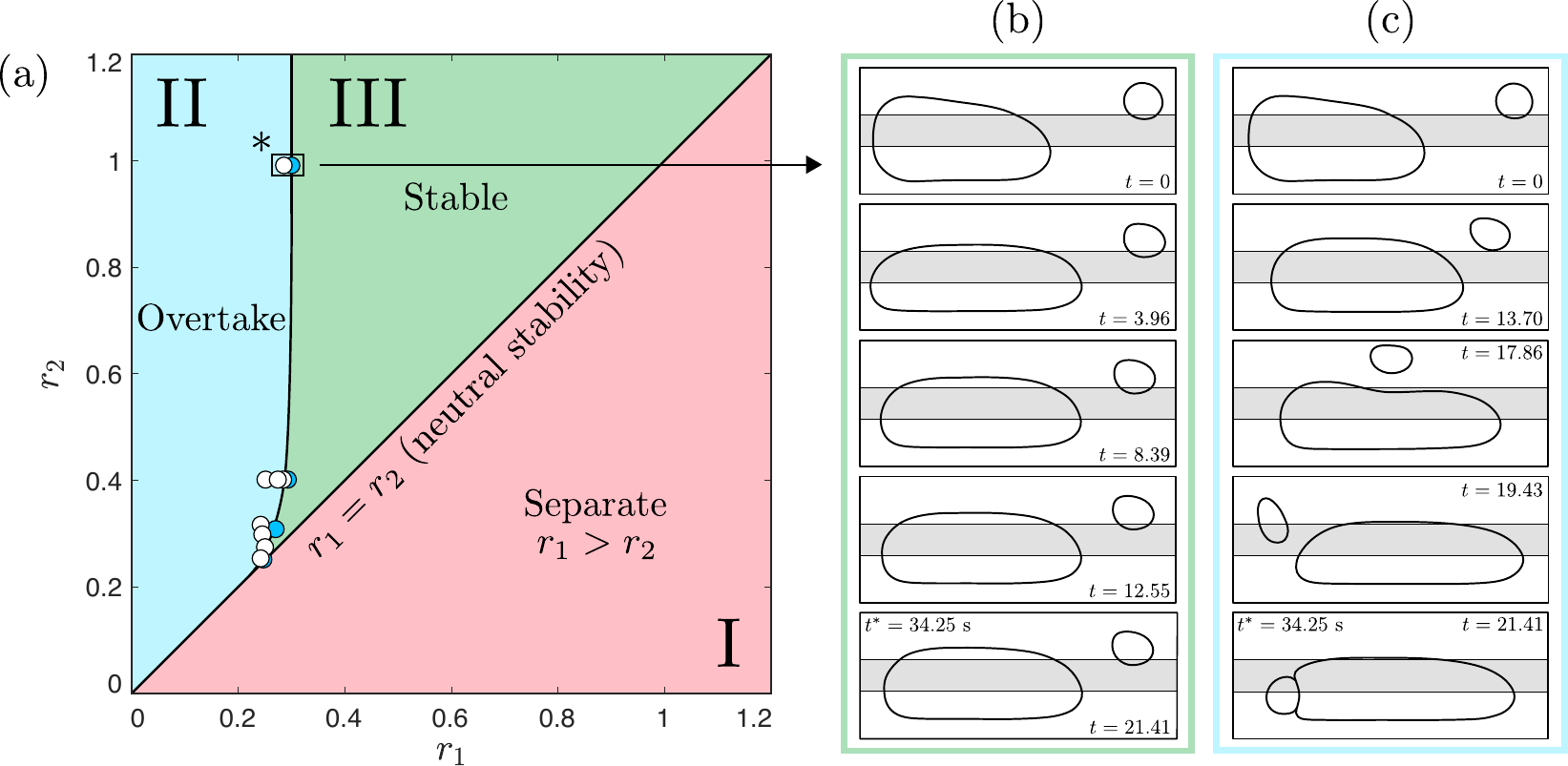}
\caption{(a) Two-dimensional phase diagram that classifies the long-term behaviour of two bubbles initialised on opposite sides of the rail when propagated from rest at $Q=0.01$. The leading bubble has size $r_1$ and the trailing bubble has size $r_2$. The solid blue markers correspond to the bubbles organising into a stable formation, whereas the solid white markers correspond to an overtaking of the leading bubble by the trailing bubble. (b--c) Experimental time-sequences of two bubbles when propagated from rest at $Q=0.01$. The trailing bubble's size is fixed at $r_2=1.00$. The leading bubble's size is $r_1=0.30$ in (b) and $r_1=0.29$ in (c).}
\label{fig:phasediagram}
\end{figure}

The two-dimensional phase diagram in Fig.~\ref{fig:phasediagram}\textcolor{blue}{(a)} classifies the long-term behaviour of all possible bubble pairs $(r_1, \: r_2)$ when propagated from rest on opposite sides of the rail at a fixed value of the dimensionless flow rate $Q=0.01$. We note that it is qualitatively representative of all the values of flow rate investigated. The bubbles were initialised sufficiently far apart in order to prevent their immediate aggregation following the imposition of flow. We find that their dynamics can be grouped into three simply connected regions, denoted by \textbf{I}, \textbf{II} and \textbf{III}, which are described as follows.

\begin{enumerate}
\renewcommand{\theenumi}{\textbf{I}}
  \item{Within this region, where $r_1 > r_2$, the trailing bubble's isolated speed is less than that of the leading bubble and, thus, the two bubbles separate indefinitely. The two bubbles ultimately propagate independently because they do not feel the flow field perturbation that is imposed by the other bubble at sufficiently large ($D > 2$) separations.}

  \renewcommand{\theenumi}{\textbf{II}}
  \item{Within this region, where $r_1 < r_2$, the two bubbles do not organise into a stable formation and, instead, the trailing bubble overtakes the leading bubble. In this case, the flow field perturbation that is imposed by the leading bubble is too small to provoke the geometric changes required in the trailing bubble to reduce its speed to that of the leading bubble.}

\renewcommand{\theenumi}{\textbf{III}}
    \item{Within this region, where $r_1 < r_2$, the two bubbles organise into a stable formation because the flow field perturbation that is imposed by the leading bubble is able to remodel the trailing bubble sufficiently and reduce its speed to that of the leading bubble. A stable two-bubble formation exists for all $0.30 \leq r_1 < r_2$. However, a stable formation only exists within a limited interval of $r_2$ for $0.24 \leq r_1 \leq 0.29$ and the size of the interval decreases as $r_1$ decreases. We were unable to resolve any stable formations for $r_1 < 0.24$.}
   
\end{enumerate}

The pair of experimental time sequences in Figs.~\ref{fig:phasediagram}\textcolor{blue}{(b)} and \ref{fig:phasediagram}\textcolor{blue}{(c)} show the dynamical transition between regions \textbf{II} and \textbf{III} for $r_2 = 1.00$ (highlighted on the phase diagram). The two bubbles organise into a stable formation in Fig.~\ref{fig:phasediagram}\textcolor{blue}{(b)} for $r_1=0.30$, whereas the trailing bubble overtakes the leading bubble in Fig.~\ref{fig:phasediagram}\textcolor{blue}{(c)} for $r_1=0.29$. In the latter case, the two bubbles aggregate and form a compound bubble whose propagation speed is equal to that of an isolated bubble of size $\sqrt{r_1^2 + r_2^2}$ \citep{LifeAndFate}. We note that the compound bubble only exists transiently because the lubrication film that separates the two bubbles drains over time and this ultimately results in their coalescence. However, a longer channel would be required to observe this in experiments.

In summary, we find that stable steadily propagating two-bubble states are possible if the bubbles are located on opposite sides of the rail and the leading bubble is smaller than the trailing bubble. However, the leading bubble must also be sufficiently large such that its local perturbation to the flow field is sufficient to cause the required deformation of the trailing bubble that reduces its speed to match that of the leading bubble.

\subsection{Stable three-bubble formations}
\label{N_bubble}
\subsubsection{Variation of bubble order}

\begin{figure}
\includegraphics[width=\textwidth, clip]{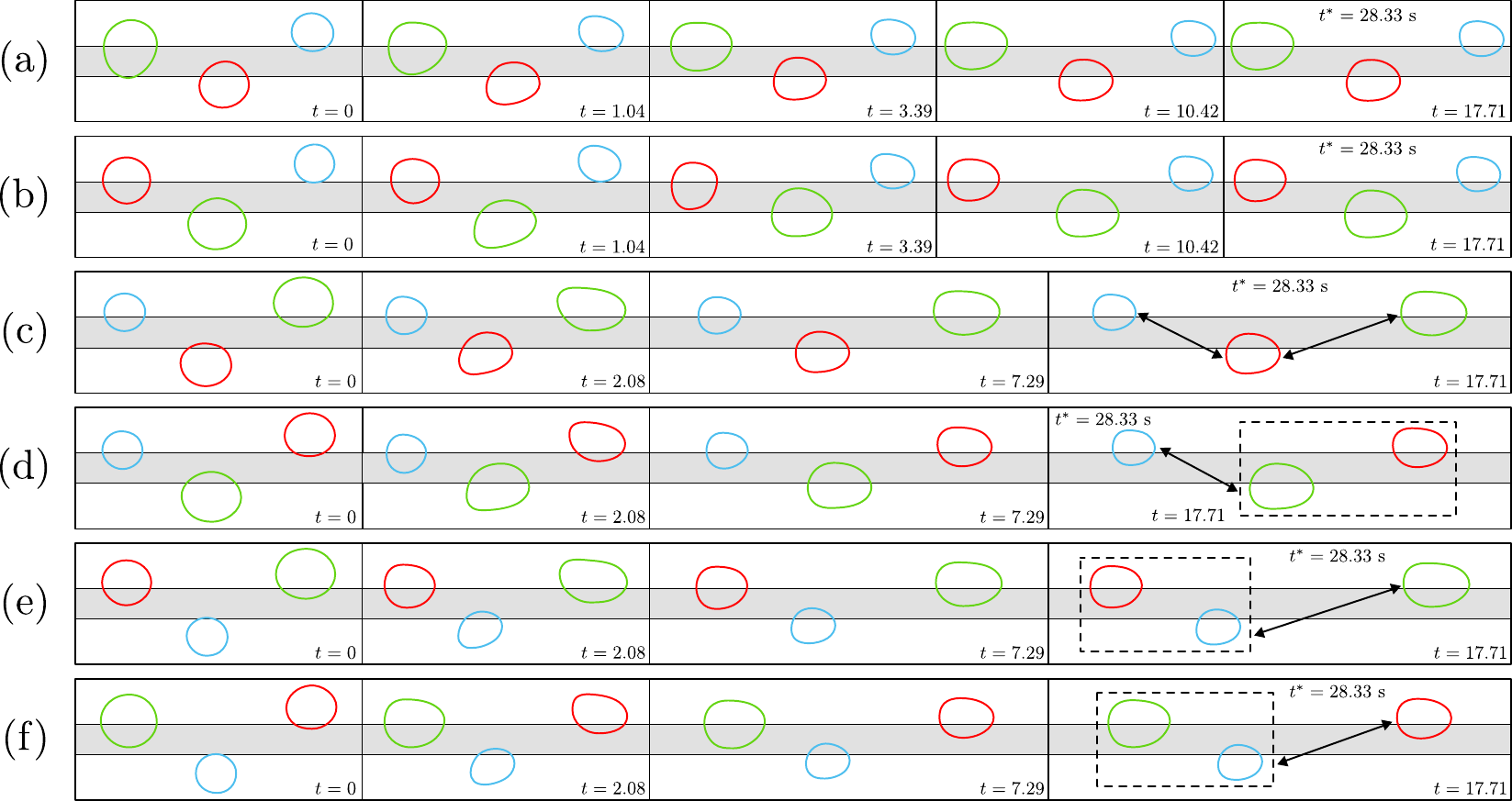}
\caption{Experimental time-sequences that show the evolution of the six distinct arrangements of three bubbles initialised in alternation on opposite sides of the rail when propagated from rest at $Q=0.01$. An arrow between a pair of bubbles in the final panels indicates that they separate indefinitely whilst a dashed box around a pair of bubbles indicates that they organise into a stable formation. The bubble sizes are $r=0.33$ (blue-coloured), $r=0.40$ (red-coloured) and $r=0.47$ (green-coloured).}
\label{fig:three_bubble_ts}
\end{figure}

\begin{figure}[b!]
\includegraphics[width=\textwidth, clip]{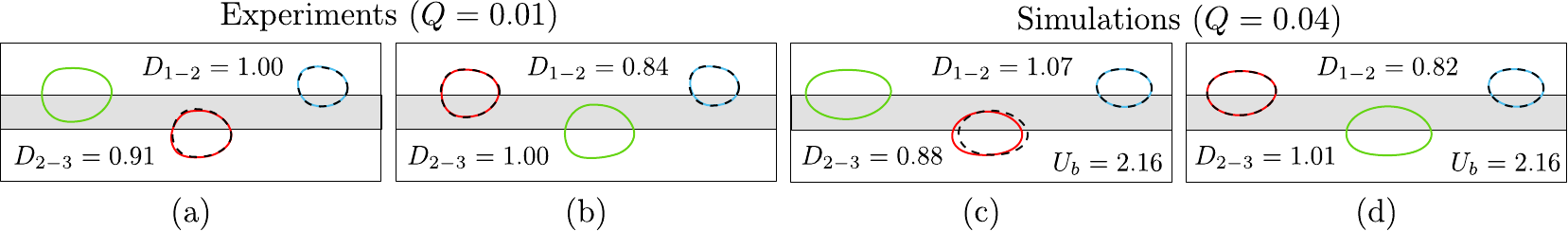}
\caption{The two stable three-bubble formations that were identified in (a, b) experiments at $Q=0.01$ and (c,~d) numerical simulations at $Q=0.04$. The bubble sizes are $r=0.33$ (blue-coloured), $r=0.40$ (red-coloured) and $r=0.47$ (green-coloured). The dashed black-coloured contours are the stable two-bubble formation bubble shapes for the blue and red-coloured bubbles.}
\label{fig:three_bubble_stable}
\end{figure} 

We will now proceed to build on our understanding of stable two-bubble formations and explore the stability of larger groups of bubbles. There are $3! = 6$ distinct initial arrangements of three bubbles lying in alternation on opposite sides of the rail and we will index the bubbles in ascending order in the direction of outlet-to-inlet. We will retain the two bubble sizes that were used in Fig.~\ref{fig:two_bubble_state} and introduce a third bubble of size $r=0.47$. The experimental time-sequences in Fig.~\ref{fig:three_bubble_ts} show the evolution of the six initial arrangements when propagated from rest at $Q = 0.01$. 

The two initial arrangements in Figs.~\ref{fig:three_bubble_ts}\textcolor{blue}{(a)} and \ref{fig:three_bubble_ts}\textcolor{blue}{(b)} organise into a stable three-bubble formation and, analogous to stable two-bubble formations, they are both led by the smallest bubble and their speeds are equal to the leading bubble's isolated speed. However, the four remaining arrangements do not organise into a stable three-bubble formation and, instead, at least one pair of neighbouring bubbles separate. For example, the three bubbles in Fig.~\ref{fig:three_bubble_ts}\textcolor{blue}{(c)} are arranged in order of decreasing size and, thus, all separate in accordance with the differences between their isolated speeds. However, in Fig.~\ref{fig:three_bubble_ts}\textcolor{blue}{(d)}, the long-term behaviour is more complex  because the leading and middle bubbles organise into a stable two-bubble formation that separates from the trailing bubble. 

We have superimposed the shapes of the two smallest bubbles in their stable two-bubble formation (dashed black-coloured contours) onto the stable three-bubble formations in Figs.~\ref{fig:three_bubble_stable}\textcolor{blue}{(a)} and \ref{fig:three_bubble_stable}\textcolor{blue}{(b)}. The leading bubble's shape does not change in both formations. The order of the two smallest bubbles is preserved in Fig.~\ref{fig:three_bubble_stable}\textcolor{blue}{(a)} and the shape of the middle bubble's front and its separation from the leading bubble's rear do not change. However, in accordance with the leading bubble's behaviour in a stable two-bubble formation, its rear inclines away from the channel's centreline because of the trailing bubble's imposed flow field perturbation. The middle and trailing bubbles are interchanged in Fig.~\ref{fig:three_bubble_stable}\textcolor{blue}{(b)} and the value of $D_{1-2}$ is equal to that which is obtained for a stable two-bubble formation for the same-sized bubbles [Fig.~\ref{fig:vary_r2}\textcolor{blue}{(b)}]. The trailing bubble retains its shape from the two-bubble formation and this confirms that a bubble's shape and position within the channel's cross-section are the predominant factors that influence its speed. Thus, because each individual bubble's speed must remain the same in order for it to remain part of the formation, the bubble shapes are only weakly influenced by their relative axial position within a stable formation.

\begin{figure}[t!]
\includegraphics[width=\textwidth, clip]{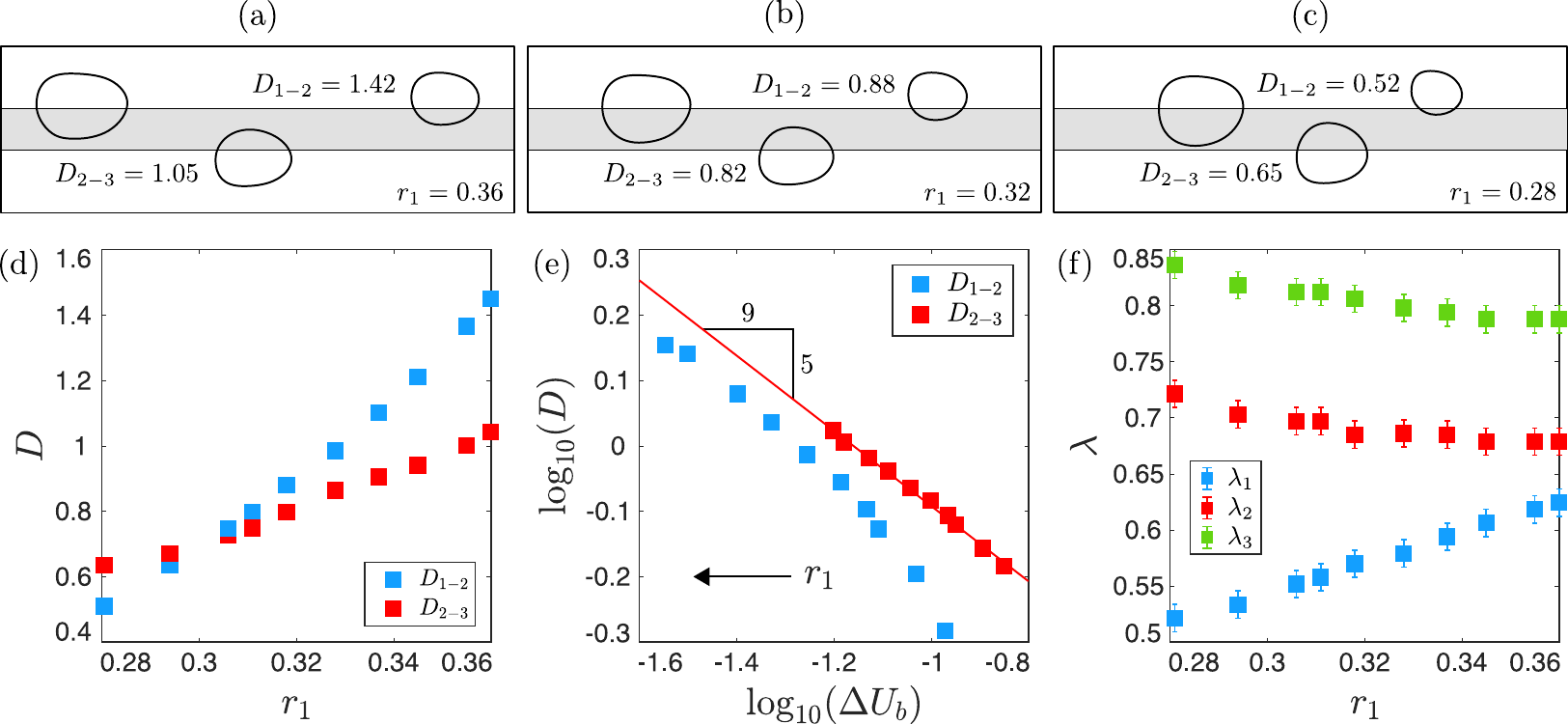}
\caption{(a--c) Experimental snapshots of various stable three-bubble formations at $Q=0.01$. The bubble sizes are $r_2=0.40$, $r_3=0.47$ and $r_1$ is varied. (d) Variation of the dimensionless separations $D_{1-2}$ and $D_{2-3}$ with $r_1$. (e) Log-log plot of $D$ versus $\Delta U_b$. (f) Variation of the three bubble widths $\lambda_1$, $\lambda_2$ and $\lambda_3$ with $r_1$. Error bars have been omitted in (d) and (e) because they are smaller than the data markers.}
\label{fig:three_bubble_distance}
\end{figure}

\begin{figure}[b!]
\includegraphics[width=\textwidth, clip]{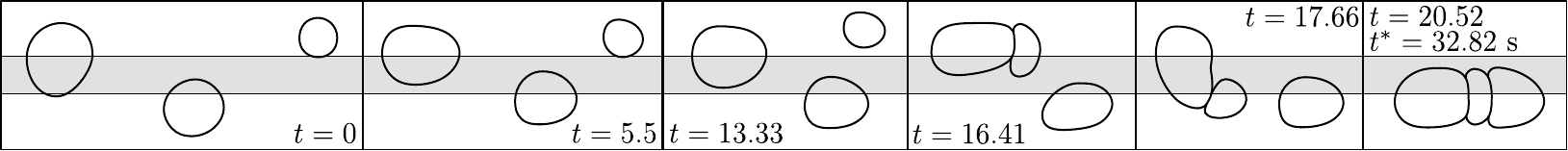}
\caption{Experimental time-sequence showing the evolution of three bubbles of size $r_1=0.26$, $r_2=0.40$ and $r_3=0.47$ when propagated from rest at $Q=0.01$. The leading bubble is too small for a stable three-bubble formation to exist.}
\label{fig:too_small}
\end{figure}

We identified qualitatively similar behaviour in numerical simulations of the depth-averaged lubrication model. The two stable formations that arise from the same bubble sizes at $Q=0.04$ are presented in Figs.~\ref{fig:three_bubble_stable}\textcolor{blue}{(c)} and \textcolor{blue}{(d)}. Once again, the propagation speed of each formation is marginally ($< 1\%$) slower than the leading bubble's isolated propagation speed. The dimensionless separation between the leading and middle bubbles increases by $0.07$ in Fig.~\ref{fig:three_bubble_stable}\textcolor{blue}{(c)} and this contrasts with experiments because it remained unchanged. However, this is the only discrepancy that was identified between experiments and numerical simulations.

\subsubsection{Variation of $D$ with $r_1$}
We have determined that the leading bubble sets the propagation speed of a stable formation of three bubbles. The middle and trailing bubbles adjust their shapes and overlap of the rail in the perturbation fields of their preceding neighbours in order to reduce their speeds to that of the leading bubble. We will now explore how a stable formation of three bubbles responds to a change in its propagation speed, which is imposed by varying the size of the leading bubble whilst the sizes of the other two bubbles and flow rate are constant. 

The experimental snapshots in Figs.~\ref{fig:three_bubble_distance}\textcolor{blue}{(a)}--\ref{fig:three_bubble_distance}\textcolor{blue}{(c)} show the stable formation as the leading bubble's size is decreased whilst the middle bubble's size $r_2=0.40$, trailing bubble's size $r_3=0.47$ and dimensionless flow rate $Q=0.01$ are fixed. The bubbles become increasingly deformed (relative to their isolated states) and the separation between each pair of neighbouring bubbles decreases [Fig.~\ref{fig:three_bubble_distance}\textcolor{blue}{(d)}]. The separation between the middle and trailing bubbles $D_{2-3}$ decreases linearly on a log-log scale as the required reduction in the trailing bubble's speed $\Delta U_{b3}$ increases.  [Fig.~\ref{fig:three_bubble_distance}\textcolor{blue}{(e)}]. This behaviour indicates that the separation between the middle and trailing bubbles is predominantly set by the required reduction in the latter's speed. However, the separation between the leading and middle bubbles $D_{1-2}$ does exhibit this linear decrease as the required reduction in the middle bubble's speed $\Delta U_{b2}$ increases because the relative changes in the leading bubble's shape are significant. The leading bubble's width decreases significantly (by approximately $20\%$) as its size is decreased [Fig.~\ref{fig:three_bubble_distance}\textcolor{blue}{(f)}]. Thus, the middle bubble has to approach the leading bubble more closely in order to feel the required flow field perturbation that results in its necessary geometric reshape.

Finally, the bubbles do not organise into a stable formation if the leading bubble is too small. For $r_1 \leq 0.26$, the leading bubble is overtaken by the middle bubble following the imposition of flow [Fig.~\ref{fig:too_small}]. The initial overtaking dynamics (i.e. in the first three panels) are qualitatively identical to the overtaking dynamics of two bubbles [Fig.~\ref{fig:phasediagram}\textcolor{blue}{(c)}] and, thus, we infer that they are uninfluenced by the trailing bubble. The three bubbles all aggregate in a complex transient evolution and create a single compound bubble. However, as discussed previously, this state only exists transiently because the lubrication films that separate the bubbles drain over time.

\begin{figure}
\includegraphics[width=\textwidth, clip]{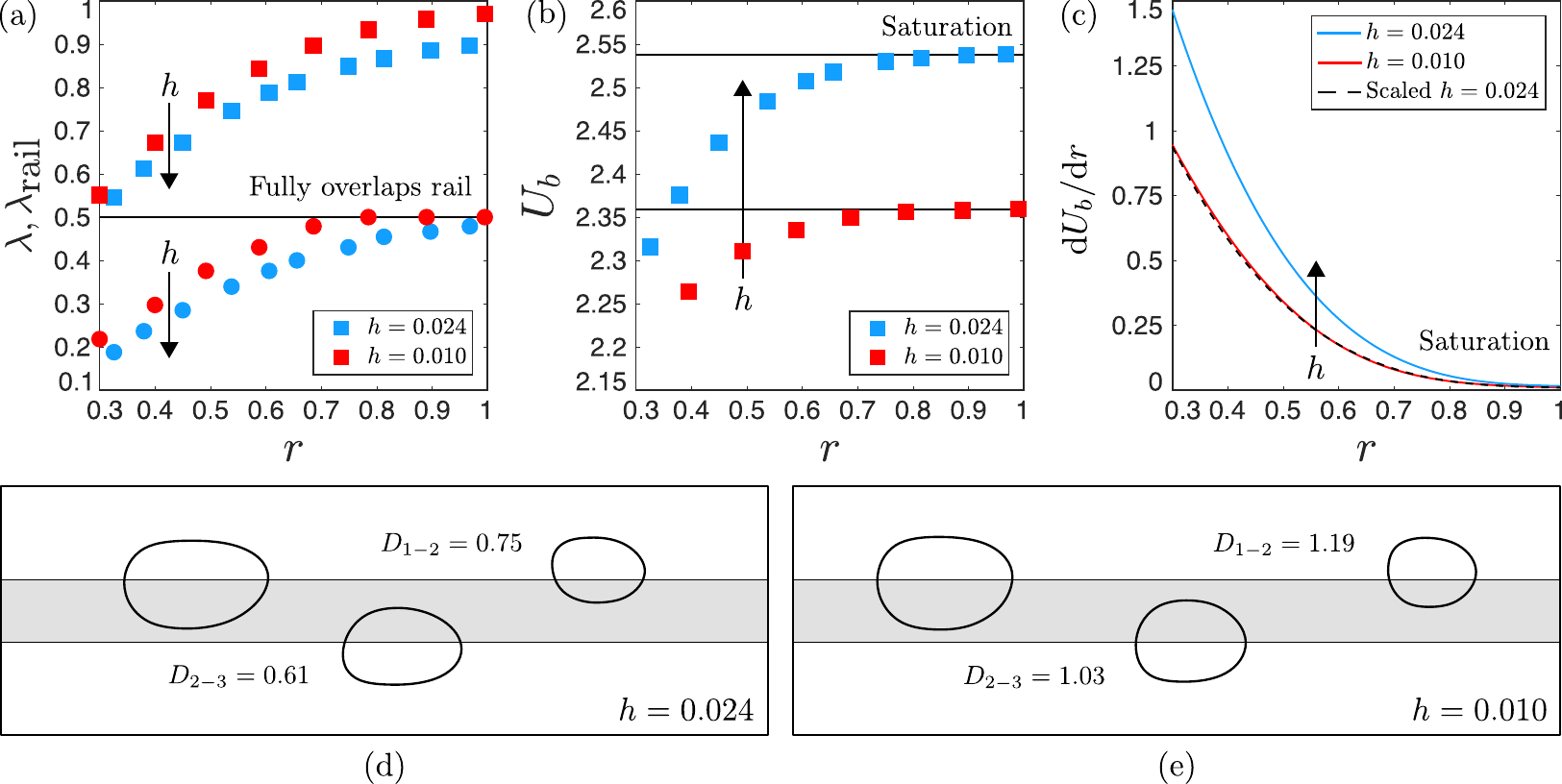}
\caption{(a) Variation of $\lambda$ and $\lambda_{\textnormal{rail}}$ with $r$ for an isolated bubble at $h=0.024$ and $h=0.010$. (b) Variation of $U_b$ with $r$ for an isolated bubble at $h=0.024$ and $h=0.010$. (c) Variation of $\mathrm{d}U_b / \mathrm{d}r$ with $r$ for an isolated bubble at $Q=0.04$; these curves were obtained by differentiating power law fits to the data points in (b). (d, e) A stable three-bubble formation at (c) $h=0.024$ and (d) $h=0.010$ in experiments. The flow rate is $Q=0.04$ and the bubble sizes are $r_1 = 0.33$, $r_2 = 0.40$ and $r_3 = 0.47$. Error bars have been omitted in all figures because they are smaller than the data markers.}
\label{fig:smaller_h_formation}
\end{figure} 

\subsubsection{Variation of $D$ with $h$}

We now turn to the influence of the rail height on the behaviour of a stable formation of bubbles. The rail height was reduced to $h = 0.010$ in the experimental channel by applying a thinner strip of PET tape to the lower glass plate. We begin by characterising the effects of the reduced rail height on the motion of isolated bubbles. The reduced rail height causes an isolated bubble to broaden, migrate closer to the channel's centreline and, thus, increase its overlap of the rail [Fig.~\ref{fig:smaller_h_formation}\textcolor{blue}{(a)}]. The geometric remodelling of an isolated bubble's shape reduces its speed towards that of an equivalent-sized bubble in a uniform channel ($U_b \approx 2$) [Fig.~\ref{fig:smaller_h_formation}\textcolor{blue}{(b)}]. In general, the difference between an isolated bubble's speed at the two rail heights increases with its size because larger bubbles occupy a greater proportion of the rail and, thus, they are increasingly affected by changes in its geometry. Furthermore, the speed of an isolated bubble becomes less-influenced by its size (i.e. the $U_b$ versus $r$ curve becomes flatter) as the rail's height is reduced. In fact, the rate of change of $U_b$ with $r$ (i.e. $\mathrm{d}U_b / \mathrm{d}r$) decreases by the scale factor $K = 0.62$ as the rail's height is reduced [Fig.~\ref{fig:smaller_h_formation}\textcolor{blue}{(e)}]. Thus, the value of $\Delta U_b$ between any two bubbles is multiplied by $K = 0.62$.

\begin{figure}
\includegraphics[width=\textwidth, clip]{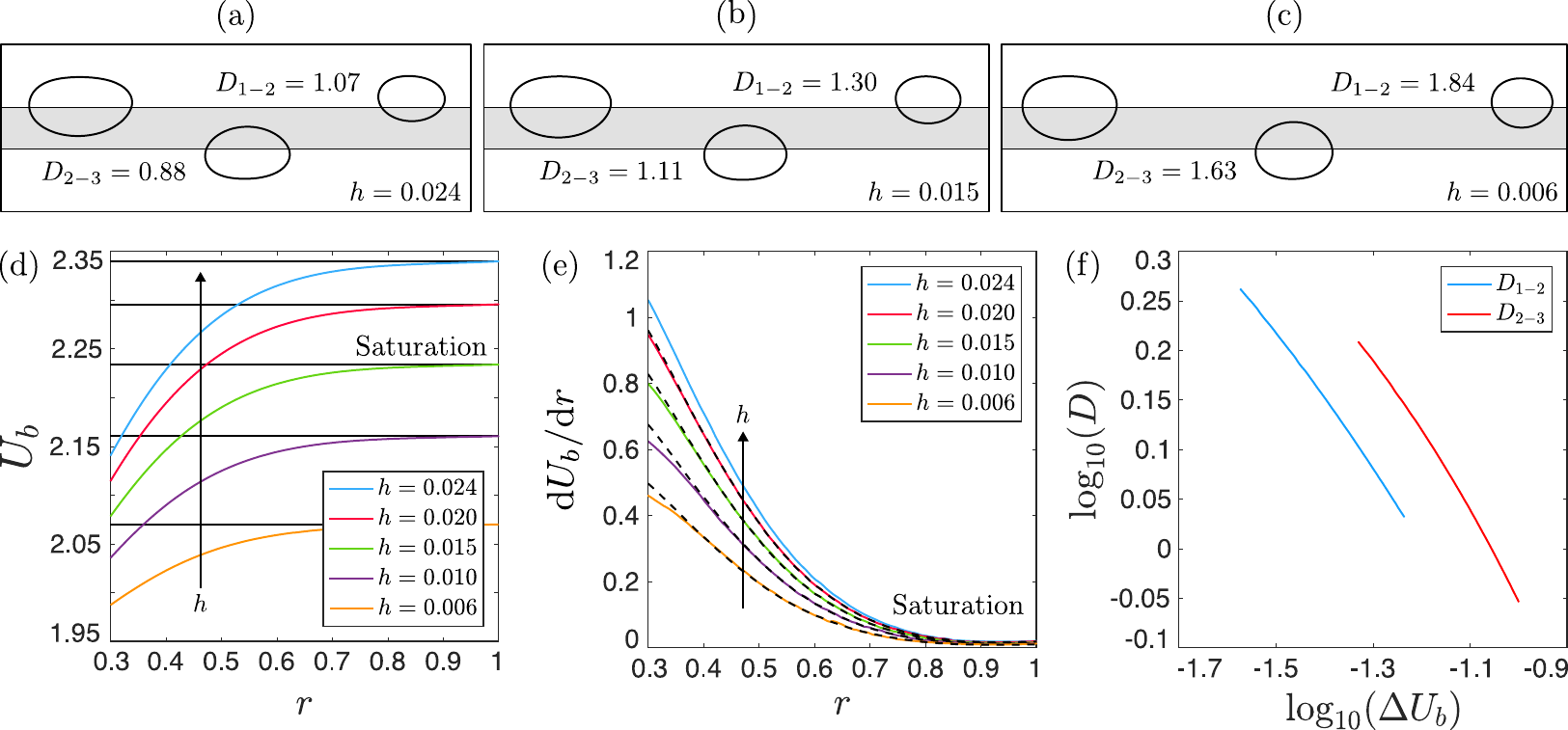}
\caption{(a--c) Stable three-bubble formations at various values of the rail height $h$ in the depth-averaged lubrication model at $Q=0.04$. The bubble sizes are $r_1=0.33$, $r_2=0.40$ and $r_3=0.47$. (d) Variation of $U_b$ with $r$ at various values of $h$. (e) Variation of $\Delta U_b / \Delta r$ with $r$ at various values of $h$. (f) Log-log plot of $D$ versus $\Delta U_b$.}
\label{fig:numerical_smaller_h}
\end{figure} 

The experimental snapshots in Figs. \ref{fig:smaller_h_formation}\textcolor{blue}{(d)} and \textcolor{blue}{(e)} show the stable formation that is formed by three bubbles of sizes $r_1 = 0.33$, $r_2 = 0.40$ and $r_3 = 0.47$ at the two rail heights. The separation between neighbouring pairs of bubbles increases as the rail's height is reduced because the required reduction in their speeds decreases. We note that $D_{1-2}$ and $D_{2-3}$ increase by approximately the same multiplicative factor and this is presumably because the bubble sizes are similar.

We find qualitatively similar behaviour in numerical simulations of the depth-averaged lubrication model. The bubbles become increasingly deformed, migrate towards the channel's centreline and the separation between neighbouring bubbles increases as the rail's height is reduced [Figs.~\ref{fig:numerical_smaller_h}\textcolor{blue}{(a)}--\ref{fig:numerical_smaller_h}\textcolor{blue}{(c)}]. The isolated bubble speeds decrease [Fig.~\ref{fig:numerical_smaller_h}\textcolor{blue}{(d)}] and $\textnormal{d} U_b / \textnormal{d} r$ decreases [Fig.~\ref{fig:numerical_smaller_h}\textcolor{blue}{(e)}] by a scale factor $K < 1$. The dashed black-coloured curves represent rescalings of the $h=0.024$ curve by $K$. The scale factor decreases as the rail's height is reduced and, thus, the isolated bubble speeds become increasingly similar. The separation between pairs of neighbouring bubbles decrease near-linearly with $\Delta U_b$ on a log-log scale [Fig.~\ref{fig:numerical_smaller_h}\textcolor{blue}{(f)}]. This behaviour indicates that the reduction in the separation between neighbouring bubbles is predominantly driven by their isolated speeds becoming increasingly similar. The small deviation from a perfect linear relationship is attributed to the accompanying changes in the bubble shapes.

\subsection{Larger bubble formations}

\subsubsection{Four-bubble formations}
\begin{figure}[t!]
\includegraphics[width=\textwidth, clip]{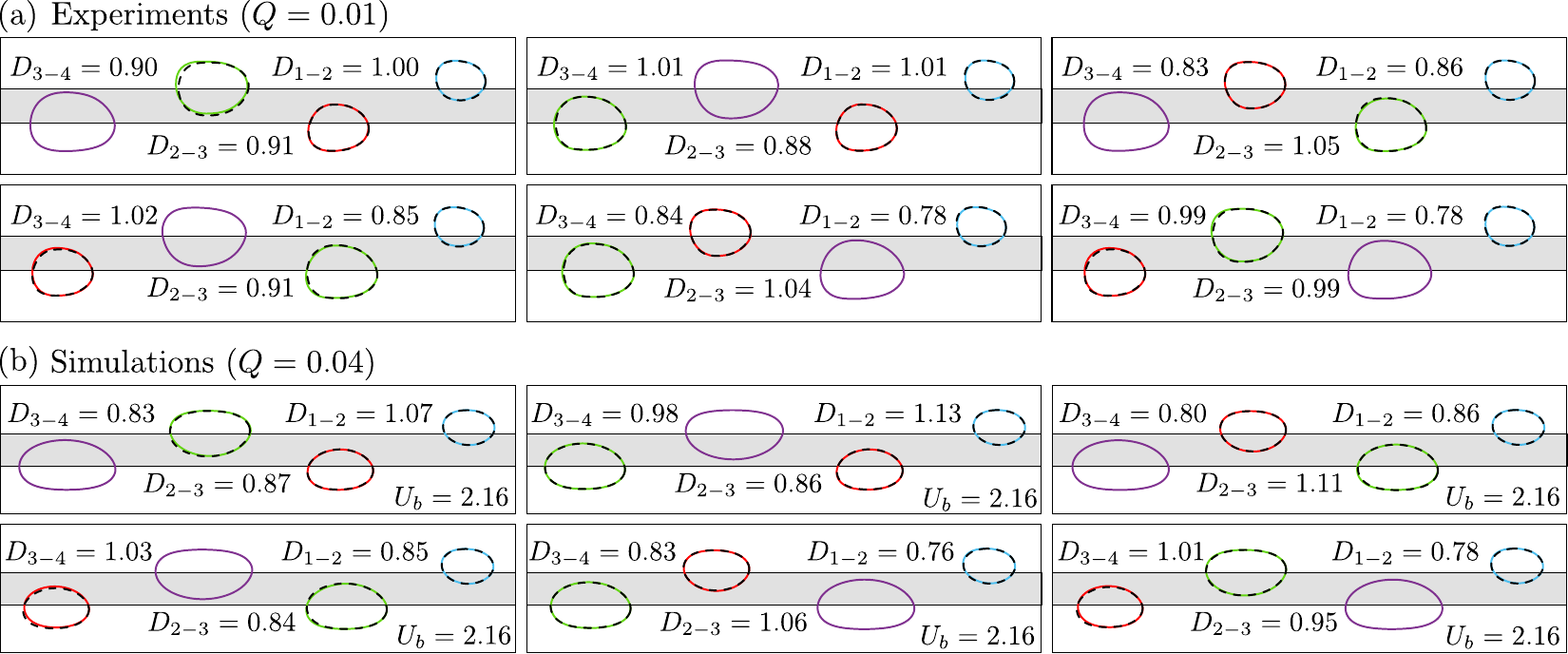}
\caption{The six stable four-bubble formations that were identified in (a) experiments at $Q = 0.01$ and (b) numerical simulations at $Q = 0.04$. The bubble sizes are $r=0.33$ (blue-coloured), $r=0.40$ (red-coloured), $r=0.47$ (green-coloured) and $r=0.54$ (purple-coloured). The dashed black-coloured contours are the stable three-bubble formation bubble shapes for the blue, red and green-coloured bubbles.}
\label{fig:four_experiments}
\end{figure}

The fundamental design principle for stable two- and three-bubble formations is that the leading bubble must be the smallest and, therefore, slowest of the bubbles. We find that the same principle applies for a group of four bubbles. There are $4! = 24$ distinct initial arrangements of four bubbles lying in alternation on opposite sides of the rail. We will retain the three bubble sizes that were used in Fig.~\ref{fig:three_bubble_ts} and introduce a fourth bubble of size $r=0.54$. Fig.~\ref{fig:four_experiments} shows the six stable formations that were identified in experiments at $Q=0.01$ and numerical simulations at $Q=0.04$. The six formations are all led by the smallest bubble and their speeds are equal to and marginally ($< 1\%$) slower than the leading bubble's isolated speed, respectively, in experiments and numerical simulations. Furthermore, the bubble shapes are approximately constant in all six formations and this, again, confirms that a bubble's speed is predominantly determined by its shape and position within the channel's cross-section. The remaining eighteen arrangements do not organise into stable formations and, instead, they partition into subgroups of independently propagating bubbles. The subgroups, depending on the order of the bubbles, are comprised of smaller stable formations and/or single bubbles.

\subsubsection{Larger bubble formations}
\begin{figure}[t!]
\includegraphics[width=\textwidth, clip]{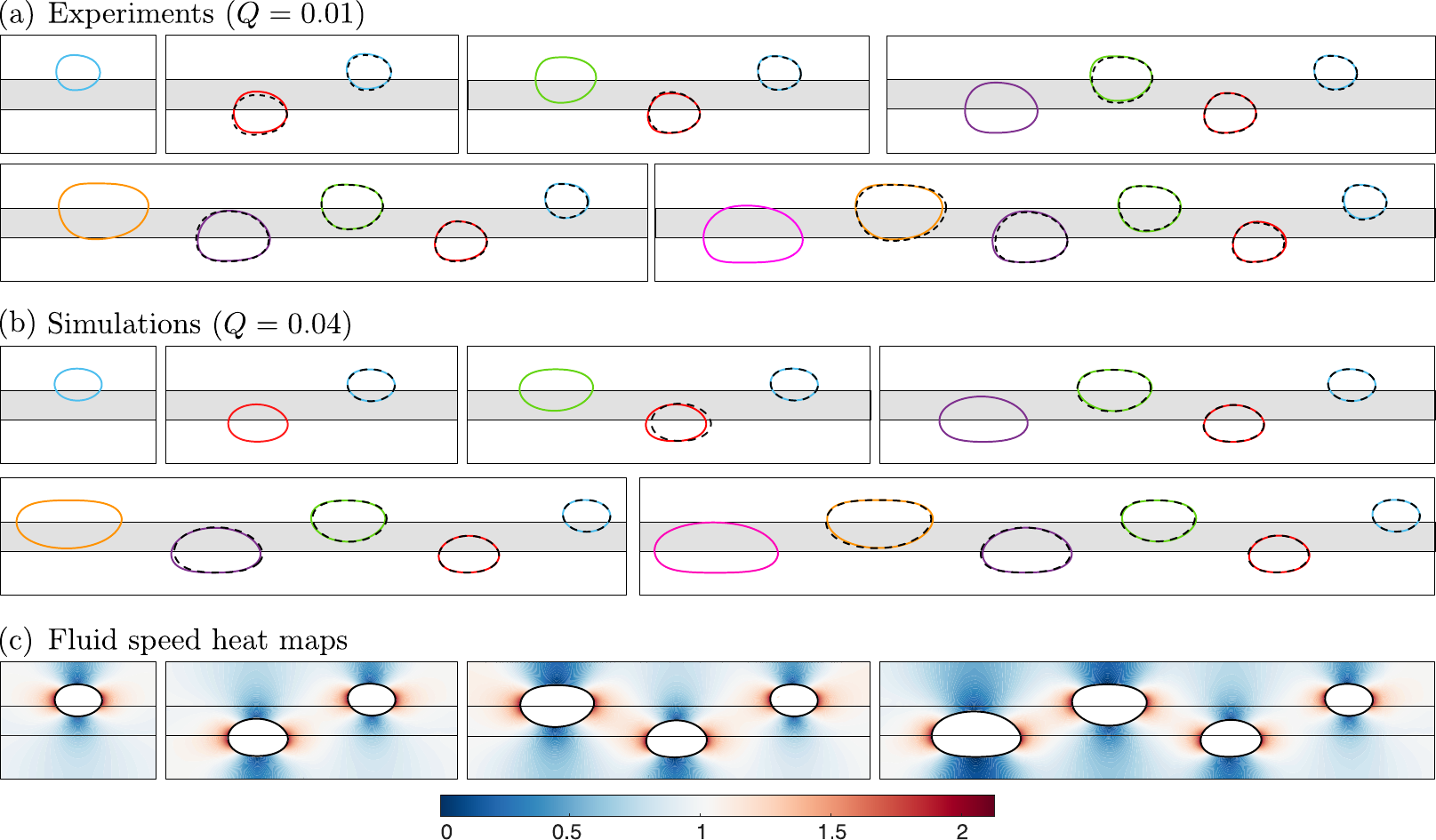}
\caption{The sequential process of building up to a stable formation of six bubbles in (a) experiments at $Q=0.01$ and (b) numerical simulations at $Q=0.04$. The previous stable formation (dashed black-coloured contours) has been superimposed onto each panel. The bubble sizes are $r=0.33$ (blue-coloured), $r=0.40$ (red-coloured), $r=0.47$ (green-coloured), $r=0.54$ (purple-coloured), $r=0.61$ (yellow-coloured) and $r=0.68$ (magenta-coloured).}
\label{fig:building_up}
\end{figure}

The behaviours that we have observed suggest that any particular bubble is primarily influenced by the flow field perturbation that is imposed by the bubble that is situated immediately ahead of it. Thus, the leading bubble propagates as if it were isolated because the flow field in the vicinity of its front is undisturbed by the flow field perturbations that arise from the trailing bubbles. The trailing bubbles must be successively slowed down by a `domino effect' of the mechanism that was described in Sec.~\ref{two_bubble} in order for there to be a stable (steadily propagating) state. The trailing bubbles adjust their shape and overlap of the rail in the perturbation fields of their preceding neighbours in order to reduce their speeds to that of the leading bubble. The local influence of each bubble's flow field perturbation means that the established design principles will remain unchanged when we consider the addition of more bubbles to the system and, thus, we infer that such formations can be extended indefinitely. We, therefore, expect that any arrangement of bubbles lying in alternation on opposite sides of the rail will organise into a stable formation provided that: (i) they are led by the smallest bubble and (ii) the leading bubble's size exceeds the minimum bubble size that is capable of slowing down the fastest (i.e. $r \geq r_s$) bubble. We have confirmed this up to six bubbles in experiments and numerical simulations; Fig.~\ref{fig:building_up} is an example of the sequential process of building up to a stable formation of six bubbles from an isolated bubble. The near-invariance of the bubble shapes is highlighted in each panel by the dashed black-coloured contours whilst the numerically computed heat maps of the fluid's speed show the adjoined perturbation fields that are formed between neighbouring pairs of bubbles.

\subsection{Bubble train dynamics}
\label{bubble_train}
\begin{figure}
\includegraphics[width=\textwidth, clip]{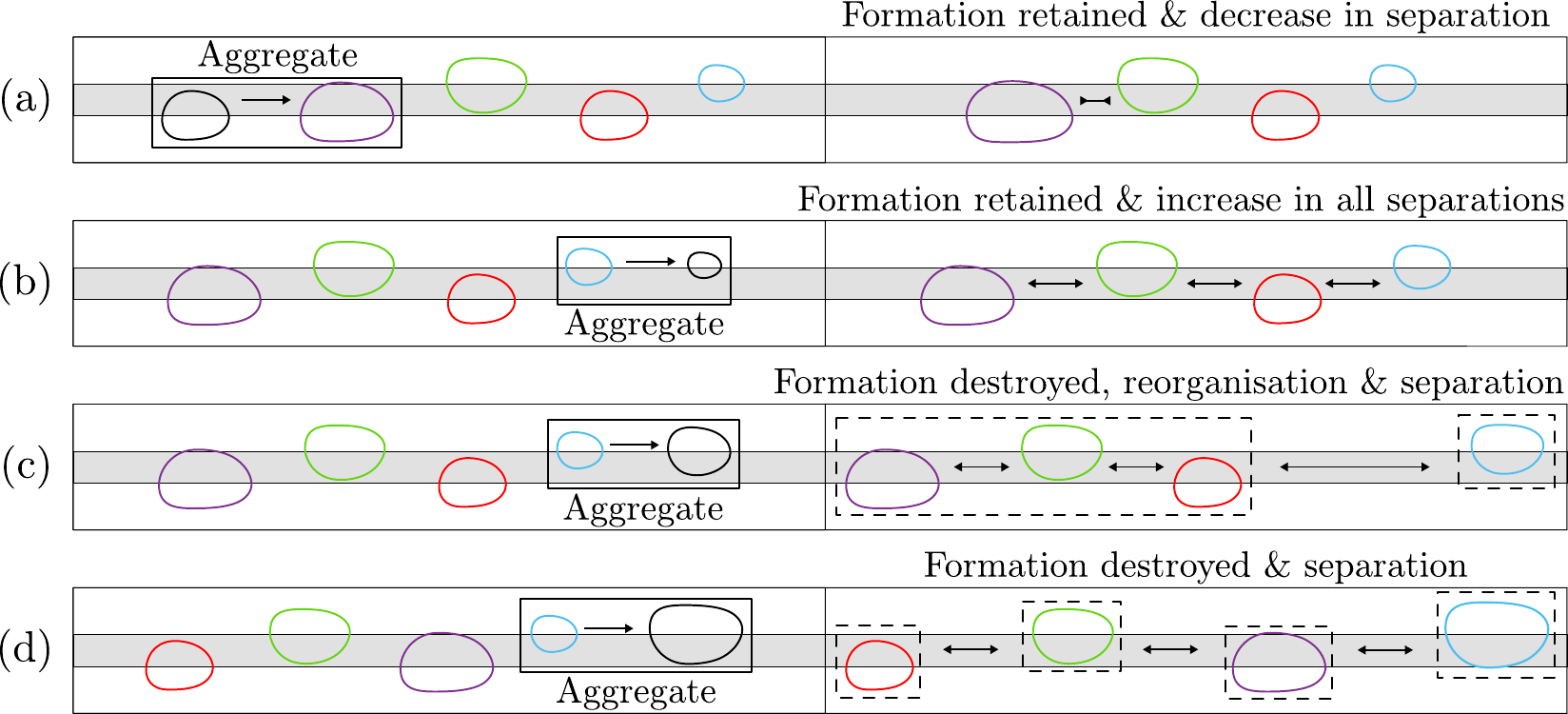}
\caption{Schematic diagrams of the reorganisation dynamics that could occur following an aggregation event involving either the leading or trailing bubbles of a stable formation. Time increases from left-to-right. (a) The size of the trailing bubble increases post-aggregation. (b--d) The size of the leading bubble increases post-aggregation. Dashed boxes in the final panels indicate independently propagating formations.}
\label{fig:conclusion_schematic}
\end{figure}

Having established the fundamental design principles of stable bubble formations, we will now proceed to discuss the expected long-term behaviour of an arbitrary train of initially well-separated bubbles whose sizes exceed the minimum bubble size that is capable of slowest down the fastest (i.e. $r \geq r_s$) bubble. We will first consider a train of $N$ bubbles lying entirely in alternation on opposite sides of the rail, for which there are two distinct behaviours.
\begin{enumerate}
    \item The bubbles will organise into a stable formation of $N$ bubbles if they are led by the smallest bubble.
    \item The bubbles will partition into $2 \leq k \leq N$ subgroups if they are not led by the smallest bubble. The subgroups, depending on the order of the bubbles, are comprised of smaller stable formations and/or single bubbles. The subgroups are arranged in order of increasing speed in the direction of inlet-to-outlet and any two consecutive subgroups will separate at a rate equal to the difference between the isolated speeds of their leading bubbles.
\end{enumerate}
The dynamics will be considerably more complex if the bubbles do not lie entirely in alternation on opposite sides of the rail because neighbouring bubbles on the same side of the rail are susceptible to aggregation events. The leading and trailing bubbles of a stable formation can aggregate with the bubble situated ahead of and behind it, respectively, based on their speeds. The bubbles will reorganise post-aggregation in accordance with the established design principles. There are several practical difficulties (e.g. an inadequate channel length) that prevented a detailed investigation of this behaviour in experiments and, similarly, numerical simulations are constrained by the finite size of the computational domain. However, based on previous results, we will proceed to infer the dynamics that would occur post-aggregation. The schematics that are provided in Fig.~\ref{fig:conclusion_schematic} act as visual aids in order to support the following descriptions of this more general case.

\begin{enumerate}

\item{The trailing bubble of a stable formation aggregates with the bubble situated behind it: the stable formation will be retained because the leading bubble remains the slowest of its constituent bubbles. However, the separation between the trailing bubble and its preceding neighbour will decrease because the former is larger and, thus, requires a greater reduction in its speed [Fig.~\ref{fig:conclusion_schematic}\textcolor{blue}{(a)}].}
\item{The leading bubble of a stable formation aggregates with the bubble situated ahead of it and it remains the smallest of the formation's bubbles post-aggregation: the stable formation will be retained because the leading bubble remains the slowest of its constituent bubbles. However, the separation between all neighbouring pairs of bubbles will increase because of the greater reductions that are required in all of their speeds [Fig.~\ref{fig:conclusion_schematic}\textcolor{blue}{(b)}].}
\item{The leading bubble of a stable formation aggregates with the bubble situated ahead of it and it does not remain the smallest of the formation's bubbles post-aggregation: the formation will be destroyed because the leading bubble does not remain the slowest of its constituent bubbles and the bubbles will reorganise in accordance with the established design principles. The two most extreme examples are those in which the remaining bubbles are arranged in order of decreasing or increasing size. In the former case, they will reorganise into a smaller formation and separate from the aggregate bubble [Fig.~\ref{fig:conclusion_schematic}\textcolor{blue}{(c)}]. In the latter case, all of the bubbles will separate [Fig.~\ref{fig:conclusion_schematic}\textcolor{blue}{(d)}]. The remaining bubbles will reorganise into a number of smaller formations in all other cases.}
\end{enumerate}
In general, an arbitrary bubble train will exhibit a combination of all such behaviours and a variety of stable bubble formations will transiently appear and disappear during the system's temporal evolution before it ultimately settles in a particular long-term state comprised of various steadily propagating subgroups that separate indefinitely.
\section{Conclusion}
\label{discussion}
We have investigated the behaviour of stable multi-bubble formations in a horizontal Hele-Shaw channel that contains an axially uniform depth reduction along its centreline. The presence of the depth reduction, or rail, allows individual bubbles to propagate at different locations within the channel's cross-section via the existence of stable asymmetric states and this can prevent the bubble aggregation that typically occurs in Hele-Shaw channels of uniform depth \citep{franco_gomez_2018}. The emergence of the multi-bubble formations is an example of complex self-organising dynamics in a confined two-phase flow and we find that there can be multiple such formations that exist for fixed system parameters. However, in contrast, groups of bubbles in a Hele-Shaw channel of uniform depth will only form multi-bubble states in the degenerate case when all of the bubbles are the same size.

The constituent bubbles of a stable formation must lie in alternation on opposite sides of the rail in order to prevent aggregation and they each propagate steadily, at the same speed, with fixed shapes. We note that \citet{Keeler2021} found two-bubble steady states in which the bubbles were placed on the same side of the rail or aligned along the centre of the channel in the depth-averaged model. However, these states were all unstable. A stable bubble formation is always led by the smallest, which is the slowest, of its constituent bubbles. The flow field ahead of the leading bubble is undisturbed by the trailing bubbles and, hence, it propagates as if it were isolated. As a consequence, the speed of a stable formation is set by the leading bubble's isolated speed. The presence of the rail allows the trailing bubbles to reduce their speeds to that of the leading bubble by adjusting their shapes and degrees of overlap of the rail within the perturbation fields of their preceding neighbours. The extent of a bubble's shape perturbation generally increases as the difference between its isolated speed and that of the leading bubble increases. However, if the leading bubble is too small ($r_1 < 0.24$), its perturbation field cannot change the shape of the trailing bubble enough to match the speed of the leading bubble and, therefore, a stable state is not possible.

The speed of each bubble is set predominantly by its shape and relative position within the channel's cross-section and each bubble's local environment is only influenced by its nearest downstream neighbour because the spatial decay of the perturbation that is imposed by each bubble is negligible beyond one channel width. The required speed of each bubble is dictated by the speed of the formation; this means that the trailing bubbles can be arranged in any order and their shapes are largely unaffected when they are interchanged. Thus, the number of possible stable arrangements for a group of bubbles increases factorially as the number of bubbles increases. The largest number of bubbles that we investigated was six because it becomes increasingly difficult to initialise larger groups of bubbles in experiments and numerical simulations are constrained by the finite size of the computational domain. However, our results suggest that stable formations will exist for arbitrarily large groups of bubbles. We have established an excellent qualitative agreement between experiments and numerical simulations of a depth-averaged lubrication model. However, a small degree of quantitative disagreement arises because the model does not incorporate the fluid films that are deposited on the upper and lower channel boundaries and this means that, in general, bubbles in the numerical simulations propagate more slowly than those in the experiments for the same flow rate. Nevertheless, the broad agreement between our two approaches indicates that the deposited fluid films do not play a fundamental role in the stabilisation of a formation of bubbles.

The fact that each bubble is influenced only by its preceding neighbour means that we can understand the general behaviour of an arbitrary train of bubbles of different sizes. The bubbles will, in general, undergo a variety of complex reorganisations when neighbouring bubbles are on the same side of the rail before they settle into a final state consisting of steadily propagating formations of bubbles that either propagate at the same speed, maintaining fixed separations, or separate indefinitely when the trailing formation propagates more slowly than the leading one. The evolution towards the final state will contain considerable structure, however, because the individual bubbles will organise into local formations that then interact. Although the bubbles might be expected to end up in approximate size order, the fact that the trailing bubbles in an individual formation can be arranged in any order means that the only definitive conclusion is that the smallest bubble will lead the most downstream formation. In contrast, in a uniform channel, the smallest bubble will always merge with any larger trailing bubbles until the final bubble in the train is the smallest remaining. In other words, the presence of the rail allows small bubbles to propagate as part of the overall train. 

The variation in the channel's depth that is provided by the rail plays two crucial roles in this system: (i) it enables simultaneous stable, steadily propagating states for individual bubbles whose centres are not aligned with the channel's centreline and (ii) it enables an extra mechanism by which bubble speeds can be changed because the volume of fluid displaced, and therefore the local viscous dissipation, can be varied without changing the bubble's shape by altering the proportion of the bubble that overlaps the rail. In the absence of the rail, the speeds of the faster bubbles cannot be altered sufficiently to form stable, steady multi-bubble states. As the height of the rail decreases, the differences between speeds of the bubbles of different sizes decreases, the bubble centres move closer to the centreline of the channel and the separation between neighbouring bubbles increases in the steady formations. We expect that the stability of the multi-bubble formations will be lost in the limit of a vanishing rail height ($h^* \rightarrow 0$) because the asymmetric, steadily propagating state is lost for individual bubbles \citep{FrancoGomez2016}. Furthermore, we also expect that their stability will be lost in the limit of the rail's height approaching that of the channel ($h^* \rightarrow H^*$) because, for sufficiently large rail heights, asymmetric bubbles no longer overlap the rail and, instead, they propagate in disconnected side-channels on opposite sides of the rail \citep{hazel_tubes}. The rail's width is also expected to have a similar influence on the stability of multi-bubble formations. In the limit of a vanishing rail width ($w^* \rightarrow 0$), the rail will tend towards a wire and impose a localised perturbation at the tip of a bubble similar to the experiments that are described by \citet{Zocchi}. Conversely, in the limit that the rail's width approaches that of the channel ($w^* \rightarrow W^*$), the bubbles will fully overlap it and propagate along the channel's centreline. However, we have not pursed a detailed investigation of these limits.

The transition from simple aggregation and separation events in a channel of uniform depth to complex dynamics involving large numbers of steadily propagating states containing multiple bubbles, simply by introducing a small perturbation to the channel's geometry, is remarkable. The presence of the additional steady states will slow down or even remove a number of bubble aggregation events and it may be, therefore, that such a small geometric variation could help to stabilise bubble trains propagating over long distances to slight variations in bubble size.

\acknowledgements{This work was supported via EPSRC grants EP/P026044/1 and EP/T008725/1 and an EPSRC DTP studentship (JL).}

\bibliography{Main_text.bib}

\end{document}